\documentclass[onecolumn]{article}%
\usepackage{amsmath}
\usepackage{amsfonts}
\usepackage{amssymb}
\usepackage{graphicx}
\usepackage{revsymb}%
\begin{document}

\title{Spin Foams for the $SO(4,C)\ $BF theory and the $SO(4,C)$ General Relativity.}
\author{Suresh. K.\ Maran}
\maketitle

\begin{abstract}
The Spin Foam Model for the $SO(4,C)\ $BF theory is discussed. The
Barrett-Crane intertwiner for the $SO(4,C)\ $general relativity is
systematically derived. The $SO(4,C)$ Barret-Crane interwiner is unique. The
propagators of the $SO(4,C)$ Barrett-Crane model are discussed. The asymptotic
limit of the $SO(4,C)$ general relativity is discussed. The asymptotic limit
is controlled by the $SO(4,C)$ Regge calculus.

\end{abstract}

\section{Introduction}

The Spin Foam model of the BF theory \cite{BF} for the gauge group $SO(4,C)$
is discussed. The Barrett-Crane model \cite{BCReimmanion} of the
$SO(4,C)\ $general relativity is systematically derived. The $SO(4,C)\ $%
Barrett-Crane model has been used to develop the concept of reality conditions
for the Barrett-Crane models \cite{RealitySpinFoam}.

The asymptotic limit of the $SO(4,C)$ general relativity is discussed. The
asymptotic limit \cite{PonzanoReggeModel}, \cite{AngMomClassLimit} is
controlled by the $SO(4,C)$ Regge calculus which unifies the Regge calculus
\cite{ReggeCalc} theories for all the real general relativity cases for the
four dimensional signatures.

\section{Spin foam of the $SO(4,C)$ BF model}

Consider a four dimensional submanifold $M$. Let $A$ be a $SO(4,C)$ connection
1-form and $B^{ij}$ a complex bivector valued $2$-form on $M$. Let $F$ be the
curvature 2-form of the connection $A$. Then I define a real continuum
BF\ theory action \cite{RealitySpinFoam},%
\begin{equation}
S_{BF}(A,B_{ij},\bar{A},\bar{B}_{ij})=\operatorname{Re}\int_{M}B\wedge F,
\label{BFaction}%
\end{equation}
where $A,B_{ij}$ and their complex conjugates are considered as independent
free variables.

The Spin foam model for the $SO(4,C)$ BF\ theory action can be derived
from\ the discretized BF action by using the path integral quantization as
illustrated in Ref:\cite{ooguriBFderv} for compact groups. Let $\Delta$ be a
simplicial manifold obtained by a triangulation of $M$. Let $g_{e}\in SO(4,C)$
be the parallel propagators associated with the edges (three-simplices)
representing the discretized connection. Let $H_{b}=%
%TCIMACRO{\tprod _{e\supset b}}%
%BeginExpansion
{\textstyle\prod_{e\supset b}}
%EndExpansion
g_{e}$ be the holonomies around the bones (two-simplices) in the four
dimensional matrix representation of $SO(4,C)$ representing the curvature. Let
$B_{b}$ be the $4\times4$ antisymmetric complex matrices corresponding to the
dual Lie algebra of $SO(4,C)$ corresponding to the discrete analog of the $B$
field. Then the discrete BF action is
\[
S_{d}=\operatorname{Re}\sum_{b\in M}tr(B_{b}\ln H_{b}),
\]
which is considered as a function of the $B_{b}$'s and $g_{e}$'s. Here $B_{b}$
the discrete analog of the $B$ field are $4\times4$ antisymmetric complex
matrices corresponding to dual Lie algebra of $SO(4,C)$. The $\ln$ maps from
the group space to the Lie algebra space. The trace is taken over the Lie
algebra indices. Then the\ quantum partition function can be calculated using
the path integral formulation as,%

\[
Z_{BF}(\Delta)=\int\prod_{b}dB_{b}d\bar{B}_{b}\exp(iS_{d})\prod_{e}dg_{e}%
\]%
\begin{equation}
=\int\prod_{b}\delta(H_{b})\prod_{e}dg_{e}, \label{eq.del}%
\end{equation}
where $dg_{e}$ is the invariant measure on the group $SO(4,C)$. The invariant
measure can be defined as the product of the bi-invariant measures on the left
and the right $SL(2,C)$ matrix components. Please see appendix A and B\ for
more details. Similar to the integral measure on the $B$'s an explicit
expression for the $dg_{e}$ involves product of conjugate measures of complex coordinates.

Now consider the identity
\begin{equation}
\delta(g)=\frac{1}{64\pi^{8}}\int d_{\omega}tr(T_{\omega}(g))d\omega,
\label{eq.del.exp}%
\end{equation}
where the $T_{\omega}(g)$ is a unitary representation of $SO(4,C),$ where
$\omega=$ $\left(  \chi_{L},\chi_{R}\right)  $ such that $n_{L}+n_{R}$ is
even, $d_{\omega}=\left\vert \chi_{L}\chi_{R}\right\vert ^{2}$. The details of
the representation theory are discussed in appendix B. The integration with
respect to $d\omega$ in the above equation is interpreted as the summation
over the discrete $n$'s and the integration over the continuous $\rho$'s.

By substituting the harmonic expansion for $\delta(g)$ into the equation
(\ref{eq.del}) we can derive the spin foam partition of the $SO(4,C)\ BF$
theory as explained in Ref:\cite{JCBintro} or Ref:\cite{ooguriBFderv}. The
partition function is defined using the $SO(4,C)$ intertwiners and the
$\{15\omega\}$ symbols.

The relevant intertwiner for the $BF\ $spin foam is%
\[
i_{e}=%
%TCIMACRO{\FRAME{itbphF}{0.518in}{0.7628in}{0.3814in}{}{}{4dintw.eps}%
%width 0.518in;  height 0.7628in;  depth 0.3814in;  original-width 2.6913in;
%original-height 3.9773in;  cropleft "0";  croptop "1";  cropright "1";
%cropbottom "0";  filename '4dintw.eps';file-properties "XNPEU";}}}%
%BeginExpansion
\raisebox{-0.3814in}{\includegraphics[
height=0.7628in,
width=0.518in
]%
{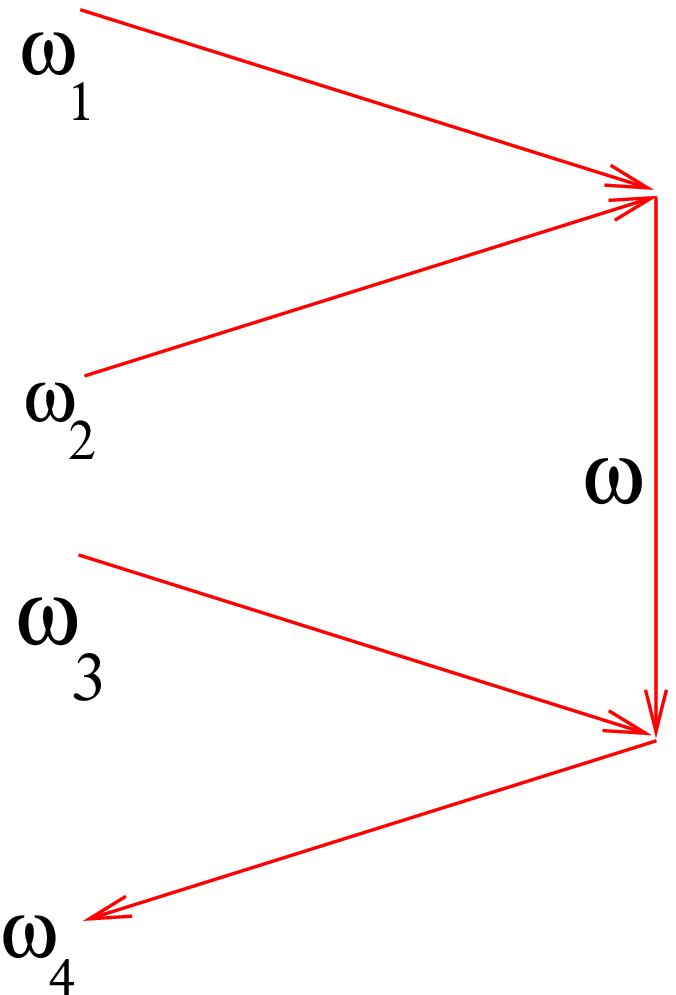}%
}%
%EndExpansion
.
\]
The nodes where the three links meet are the Clebsch-Gordan coefficients of
$SO(4,C)$. The Clebsch-Gordan coefficients of $SO(4,C)$ are just the product
of the Clebsch-Gordan coefficients of the left and the right handed $SL(2,C)$
components. The Clebsch-Gordan coefficients of $SL(2,C)$ are discussed in the
references \cite{IMG} and \cite{NaimarckClebsch}.

The quantum amplitude associated with each simplex $s$ is given below and can
be referred to as the $\{15\omega\}$ symbol,%

\[
\{15\omega\}=%
%TCIMACRO{\FRAME{itbphF}{1.6016in}{1.548in}{0.8026in}{}{}{15w.eps}%
%width 1.6016in;  height 1.548in;  depth 0.8026in;  original-width 5.5011in;
%original-height 5.316in;  cropleft "0";  croptop "1";  cropright "1";
%cropbottom "0";  filename '15w.eps';file-properties "XNPEU";}}}%
%BeginExpansion
\raisebox{-0.8026in}{\includegraphics[
height=1.548in,
width=1.6016in
]%
{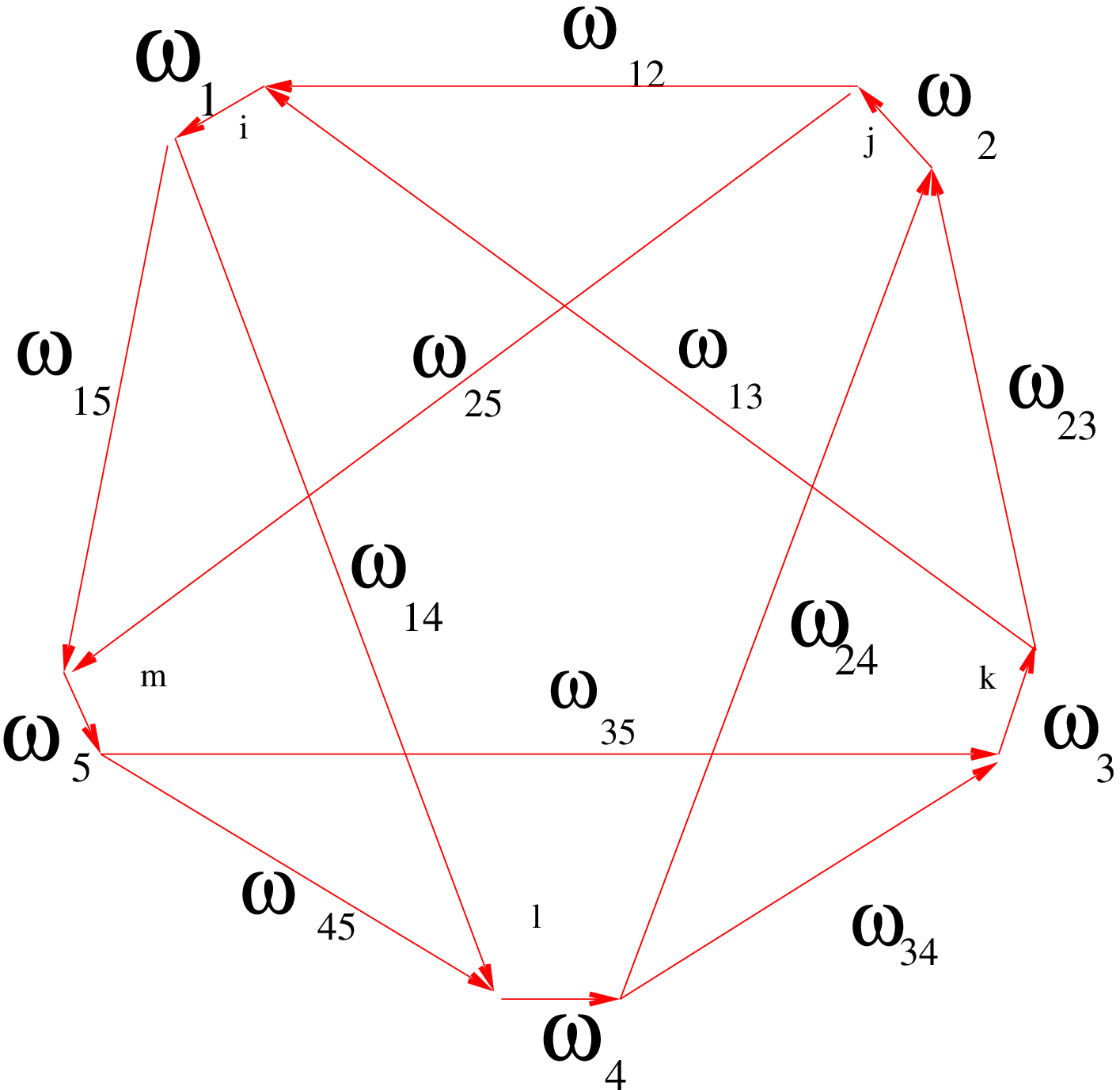}%
}%
%EndExpansion
.
\]
The final partition function is
\begin{equation}
Z_{BF}(\Delta)=\int\limits_{\{\omega_{b,}\omega_{e}\}}\prod_{b}\frac
{d_{\omega_{b}}}{64\pi^{8}}\prod_{s}Z_{BF}(s)\prod_{b}d\omega_{b}\prod
_{e}d\omega_{e}, \label{eq.6}%
\end{equation}
where the $Z_{BF}(s)=\{15\omega\}$ is the amplitude for a four-simplex $s$.
The $d_{\omega_{b}}=\left\vert \chi_{L}\chi_{R}\right\vert ^{2}$ term is the
quantum amplitude associated with the bone $b$.\ Here $\omega_{e}$ is the
internal representation used to define the intertwiners. Usually $\omega_{e}$
is replaced by $i_{e}$ to indicate the intertwiner. The set $\left\{
\boldsymbol{\omega}_{b,}\boldsymbol{\omega}_{e}\right\}  $ of all $\omega_{b}%
$'s and $\omega_{e}$'s is usually called a \textbf{coloring} of the bones and
the edges. This partition function may not be finite in general.

It is well known that the$\ BF$ theories are topological field theories. A
priori one cannot expect this to be true for the case of the BF spin foam
models because of the discretization of the BF action. For the spin foam
models of the BF\ theories for compact groups, it has been shown that the
partition functions are triangulation independent up to a factor
\cite{SpinFoamDiag}. This analysis is purely based on spin foam diagrammatics
and is independent of the group used as long the BF spin foam is defined
formally by equation (\ref{eq.del}) and the harmonic expansion in equation
(\ref{eq.del.exp}) is formally valid. So one can apply the spin foam
diagrammatics analysis directly to the $SO(4,C)$ BF spin foam and write down
the triangulation independent partition function as%

\[
Z_{BF}^{^{\prime}}(\Delta)=\tau^{n_{4}-n_{3}}Z_{BF}(\Delta)
\]
using the result from \cite{SpinFoamDiag}. In the above equation $n_{4},n_{3}$
is number of four bubbles and three bubbles in the triangulation $\Delta$ and
\begin{align*}
\tau &  =\delta_{SO(4,C)}(I)\\
&  =\frac{1}{64\pi^{8}}\int d_{\omega}^{2}d\omega.
\end{align*}
The above integral is divergent and so the partition functions need not be
finite. The normalized partition function is to be considered as the proper
partition function because the BF\ theory is supposed to be topological and so
triangulation independent.

\section{The $SO(4,C)$ Barrett-Crane Model}

\subsection{Classical $SO(4,C)$ General Relativity}

Consider a four dimensional manifold $M$. Let $A$ be a $SO(4,C)$ connection
1-form and $B^{ij}$ be a complex bivector valued $2$-form on $M$. I\ would
like to restrict myself to the non-degenerate general relativity in this
section by assuming $b=\frac{1}{4}\epsilon^{abcd}B_{ab}\wedge B_{cd}\neq0$.
The Plebanski action for the $SO(4,C)$ general relativity is obtained by
adding a Lagrange multiplier term to impose the Plebanski constraint to the BF
theory action given in equation :(\ref{BFaction}). A\ simple way of writing
the action \cite{MPR1} is%
\begin{equation}
S_{C}(A,B_{ij},\bar{A},\bar{B}_{ij},\phi)=\operatorname{Re}\left[  \int
_{M}tr(B\wedge F)+\frac{b}{2}\phi^{abcd}B_{ab}\wedge B_{cd}\right]  \text{,}
\label{GRactionComplex}%
\end{equation}
where $\phi$ is a complex tensor with the symmetries of the Riemann curvature
tensor such that $\phi^{abcd}\epsilon_{abcd}=0$. The physics corresponding to
the extrema of the above action has been discussed by me in Ref:
\cite{RealitySpinFoam}. Two important results are

\begin{itemize}
\item The Plebanski constraint imposes the condition $B_{ab}^{ij}=$
$\theta_{a}^{[i}\theta_{b}^{j]}$ where $\theta_{a}^{i}$ is a complex tetrad
field \cite{Plebanski}, \cite{ClassicalAreaReal}.

\item The field equations correspond to the $SO(4,C)$ general relativity on
the manifold $M$ \cite{ClassicalAreaReal}.
\end{itemize}

\subsubsection{Relation to Complex Geometry}

Let $M$ be a real analytic manifold. Let $M_{c}$ be the complex analytic
manifold which is obtained by analytically continuing the real coordinates on
$M$. The analytical continuation of the field equations and their solutions on
$M$ to complex $M_{c}$ can be used to define complex general relativity.
Conversely, the field equations of complex general relativity or their
solutions on $M_{c}$ when restricted to $M$ defines the $SO(4,C)$ general
relativity. Because of these properties the action $S$ can also be considered
as an action for complex general relativity.

Now consider the relation between the complex general relativity on $M_{c}$
and\ the $SO(4,C)$ general relativity on $M$. This relation critically depends
on $M$ being a real analytic manifold. It also depends on the fields on it
being analytic on some region may be except for some singularities. If the
fields and the field equations are discretized we lose the relation to complex
general relativity. Thus it is also not meaningful to relate a $SO(4,C)$
Barrett-Crane Model to complex general relativity. If the $SO(4,C)$
Barrett-Crane model has a semiclassical continuum general relativity limit
then a relation to complex general relativity may be recovered.

\subsection{The $SO(4,C)$ Barrett-Crane Constraints}

My goal here is to systematically construct the Barrett-Crane model of the
$SO(4,C)$ general relativity. In the previous section I\ discussed the
$SO(4,C)$ BF spin foam model. The basic elements of the BF spin foams are spin
networks built on graphs dual to the triangulations of the four simplices with
arbitrary intertwiners and the principal unitary representations of $SO(4,C)$
discussed in appendix B. These closed spin networks can be considered as
quantum states of four simplices in the BF\ theory and the essence of these
spin networks is mainly gauge invariance. To construct a spin foam model of
general relativity these spin networks need to be modified to include the
Plebanski Constraints in the discrete form.

A\ quantization of a four-simplex for the Riemannian general relativity was
proposed by Barrett and Crane \cite{BCReimmanion}. The bivectors $B_{i}$
associated with the ten triangles of a four-simplex in a flat Riemannian space
satisfy the following properties called the Barrett-Crane
constraints\footnote{I\ would like to refer the readers to the original paper
\cite{BCReimmanion} for more details.}:

\begin{enumerate}
\item The bivector changes sign if the orientation of the triangle is changed.

\item Each bivector is simple.

\item If two triangles share a common edge, then the sum of the bivectors is
also simple.

\item The sum of the bivectors corresponding to the edges of any tetrahedron
is zero. This sum is calculated taking into account the orientations of the
bivectors with respect to the tetrahedron.

\item The six bivectors of a four-simplex sharing the same vertex are linearly independent.

\item The volume of a tetrahedron calculated from the bivectors is real and non-zero.
\end{enumerate}

The items two and three can be summarized as follows:
\[
B_{i}\wedge B_{j}=0~\forall i,j,
\]
where $A\wedge B=\varepsilon_{IJKL}A^{IJ}B^{KL}$ and the $i,j$ represents the
triangles of a tetrahedron. If $i=j$, it is referred to as the simplicity
constraint. If $i\neq j$ it is referred as the cross-simplicity constraints.

Barrett and Crane have shown that these constraints are sufficient to restrict
a general set of ten bivectors $E_{b}$ so that they correspond to the
triangles of a geometric four-simplex up to translations and rotations in a
four dimensional flat Riemannian space.

The Barrett-Crane constraints theory can be trivially extended to the
$SO(4,C)$ general relativity. In this case the bivectors are complex and so
the volume calculated for the sixth constraint is complex. So we need to relax
the condition of the reality of the volume.

A quantum four-simplex for Riemannian general relativity is defined by
quantizing the Barrett-Crane constraints \cite{BCReimmanion}. The bivectors
$B_{i}$ are promoted to the Lie operators $\hat{B}_{i}$ on the representation
space of the relevant group and the Barrett-Crane constraints are imposed at
the quantum level. A\ four-simplex has been quantized and studied in the case
of the Riemannian general relativity before \cite{BCReimmanion}. All the first
four constraints have been rigorously implemented in this case. The last two
constraints are inequalities and they are difficult to impose. This could be
related to the fact that the Riemannian Barrett-Crane model reveal the
presence of degenerate sectors \cite{BaezEtalAsym}, \cite{JWBCS} in the
asymptotic limit \cite{JWBRW} of the model. For these reasons here after
I\ would like to refer to a spin foam model that satisfies only the first four
constraints as an \textit{essential Barrett-Crane model}, While a spin foam
model that satisfies all the six constraints as a \textit{rigorous
Barrett-Crane model}.

Here I\ would like to derive the essential $SO(4,C)$ Barrett-Crane model. For
this one must deal with complex bivectors instead of real bivectors. The
procedure that I\ would like to use to solve the constraints can be carried
over directly to the Riemannian Barrett-Crane\ model. This derivation
essentially makes the derivation of the Barrett-Crane intertwiners for the
real and the complex Riemannian general relativity more rigorous.

\subsubsection{The Simplicity Constraint}

The group $SO(4,C)$ is locally isomorphic to $\frac{SL(2,C)\times
SL(2,C)}{Z_{2}}$. An element $B$ of the Lie algebra space of $SO(4,C)$ can be
split into the left and the right handed $SL(2,C)$ components,%
\begin{equation}
B=B_{L}+B_{R}.
\end{equation}
There are two Casimir operators for $SO(4,C)$ which are $\varepsilon
_{IJKL}B^{IJ}B^{KL}$ and $\eta_{IK}\eta_{JL}B^{IJ}B^{KL}$, where $\eta_{IK}$
is the flat Euclidean metric. In terms of the left and right handed split I
can expand the Casimir operators as%
\[
\varepsilon_{IJKL}B^{IJ}B^{KL}=B_{L}\cdot B_{L}-B_{R}\cdot B_{R}~~\text{and}%
\]%
\[
\eta_{IK}\eta_{JL}B^{IJ}B^{KL}=B_{L}\cdot B_{L}+B_{R}\cdot B_{R},
\]
where the dot products are the trace in the $SL(2,C)$ Lie algebra coordinates.

The bivectors are to be quantized by promoting the Lie algebra vectors to Lie
operators on the unitary representation space of $SO(4,C)\approx$
$\frac{SL(2,C)\times SL(2,C)}{Z_{2}}$. The relevant unitary representations of
$SO(4,C)\simeq$ $SL(2,C)\otimes SL(2,C)/Z_{2}$ are labeled by a pair
($\chi_{L},$ $\chi_{R}$) such that $n_{L}+n_{R}$ is even (appendix B). The
elements of the representation space $D_{\chi_{L}}\otimes$ $D_{\chi_{R}}$ are
the eigen states of the Casimirs and on them the operators reduce to the
following:
\begin{equation}
\varepsilon_{IJKL}\hat{B}^{IJ}\hat{B}^{KL}=\frac{\chi_{L}^{2}-\chi_{R}^{2}}%
{2}\hat{I}~~\text{and} \label{eq.1}%
\end{equation}%
\begin{equation}
\eta_{IK}\eta_{JL}\hat{B}^{IJ}\hat{B}^{KL}=\frac{\chi_{L}^{2}+\chi_{R}^{2}%
-2}{2}\hat{I}. \label{eq.2}%
\end{equation}
The equation (\ref{eq.1}) implies that on $D_{\chi_{L}}\otimes$ $D_{\chi_{R}}$
the simplicity constraint $B\wedge B=0$ is equivalent to the condition
$\chi_{L}=\pm\chi_{R}$. I would like to find a representation space on which
the representations of $SO(4,C)$ are restricted precisely by $\chi_{L}$ $=$
$\pm\chi_{R}$. Since a $\chi$ representation is equivalent to $-\chi$
representations \cite{IMG}, $\chi_{L}=+\chi_{R}$ case is equivalent to
$\chi_{L}=-\chi_{R}$ \cite{IMG}.

Consider a square integrable function $f$ $(x)$ on the complex sphere $CS^{3}$
defined by%

\[
x\cdot x=1,\forall x\in\boldsymbol{C}^{4}.
\]
It can be Fourier expanded in the representation matrices of $SL(2,C)$ using
the isomorphism $CS^{3}\simeq$ $SL(2,C)$,%
\begin{equation}
f(x)=\frac{1}{8\pi^{4}}\int Tr(F(\chi)T_{\chi}(\mathfrak{g}(x)^{-1})\chi
\bar{\chi}d\chi, \label{Cs3Exp}%
\end{equation}
where the isomorphism $\mathfrak{g}\mathrm{:}CS^{3}$ $\longrightarrow SL(2,C)$
is defined in the appendix A. The group action of $g=(g_{L},g_{R})\in SO(4,C)$
on $x$ $\in CS^{3}$ is given by
\begin{equation}
\mathfrak{g}(gx)=g_{L}^{-1}\mathfrak{g}(x)g_{R}. \label{Cs3Action}%
\end{equation}
Using equation (\ref{Cs3Exp}) I can consider the $T_{\chi}(\mathfrak{g}%
(x))(z_{1},z_{2})$ as the basis functions of $L^{2}$ functions on $CS^{3}.$
The matrix elements of the action of $g$ on $CS^{3}$ is given by (appendix B)%
\[
\int\bar{T}_{\acute{\chi}}(\mathfrak{g}(x))(\acute{z}_{1},\acute{z}%
_{2})T_{\chi}(\mathfrak{g}(gx))(z_{1},z_{2})dx=T_{-\acute{\chi}}(g_{L}%
)(\acute{z}_{1},z_{1})T_{\chi}(g_{R})(\acute{z}_{2},z_{2})\delta(\acute{\chi
}-\chi).
\]
I see that the representation matrices are precisely those of $SO(4,C)$ only
restricted by the constraint $\chi_{L}=-\chi_{R}\approx\chi_{R}$. So the
simplicity constraint effectively reduces the Hilbert space $H$ to the space
of $L^{2}$ functions on $CS^{3}$. In Ref:\cite{BFfoamHighD} the analogous
result has been shown for $SO(N,R)$ where the Hilbert space is reduced to the
space of the $L^{2}$ functions on $S^{N-1}$.

\subsubsection{The Cross-simplicity Constraints}

Next let me quantize the cross-simplicity constraint part of the Barrett-Crane
constraint. Consider the quantum state space associated with a pair of
triangles $1$ and $2$ of a tetrahedron. A general quantum state that just
satisfies the simplicity constraints $B_{1}\wedge B_{1}=0$ and $B_{2}\wedge
B_{2}=0$ is of the form $f(x_{1},x_{2})$ $\in L^{2}(CS^{3}\ast CS^{3})$,
$x_{1},x_{2}\in CS^{3}$.

On the elements of $L^{2}(CS^{3}\ast CS^{3})$ the action $B_{1}\wedge B_{2}$
is equivalent to the action of $\left(  B_{1}+B_{2}\right)  \wedge\left(
B_{1}+B_{2}\right)  $\footnote{Please notice that
\[
\left(  \hat{B}_{1}+\hat{B}_{2}\right)  \wedge\left(  \hat{B}_{1}+\hat{B}%
_{2}\right)  =\hat{B}_{1}\wedge\hat{B}_{1}+\hat{B}_{2}\wedge\hat{B}_{2}%
+2B_{1}\wedge\hat{B}_{2}.
\]
But since $\hat{B}_{1}\wedge\hat{B}_{1}=\hat{B}_{2}\wedge\hat{B}_{2}=0$ on
$f(x_{1},x_{2})$ we have
\[
\left(  \hat{B}_{1}+\hat{B}_{2}\right)  \wedge\left(  \hat{B}_{1}+\hat{B}%
_{2}\right)  f(x_{1},x_{2})=\hat{B}_{1}\wedge\hat{B}_{2}f(x_{1},x_{2}).
\]
}. This implies that the cross-simplicity constraint $B_{1}\wedge B_{2}=0$
requires the simultaneous rotation of $x_{1},x_{2}$ involve only the $\chi
_{L}$ $=$ $\pm\chi_{R}$ representations. The simultaneous action of
$g=(g_{L},g_{R})$ on the arguments of $f(x_{1},x_{2})$ is%
\begin{equation}
gf(x_{1},x_{2})=f(g_{L}^{-1}x_{1}g_{R},g_{L}^{-1}x_{2}g_{R}). \label{action}%
\end{equation}
The harmonic expansion of $f(x_{1},x_{2})$ in terms of the basis function
$T_{\chi}(\mathfrak{g}(x))(z_{1},z_{2})$ is%
\[
f(x_{1},x_{2})=F_{z_{1}z_{2}\chi_{1}\chi_{2}}^{\acute{z}_{1}\acute{z}_{2}%
}T_{\acute{z}_{1}\chi_{1}}^{z_{1}}(\mathfrak{g}(x_{1}))T_{\acute{z}_{2}%
\chi_{2}}^{z_{2}}(\mathfrak{g}(x_{2})),
\]
where I have assumed all the repeated indices are either integrated or summed
over for equation only. The rest of the calculations can be understood
graphically. The last equation can be graphically written as follows:%

\[
f(x_{1},x_{2})=\iint\limits_{\chi_{1}\chi_{2}}%
%TCIMACRO{\FRAME{itbphF}{1.3318in}{0.9193in}{0.5613in}{}{\Qlb{2}}%
%{xsimplicity1.eps}{\special{ language "Scientific Word";  type "GRAPHIC";
%width 1.3318in;  height 0.9193in;  depth 0.5613in;  original-width 3.4039in;
%original-height 2.341in;  cropleft "0";  croptop "1";  cropright "1";
%cropbottom "0";  filename 'Xsimplicity1.eps';file-properties "XNPEU";}}}%
%BeginExpansion
\raisebox{-0.5613in}{\includegraphics[
height=0.9193in,
width=1.3318in
]%
{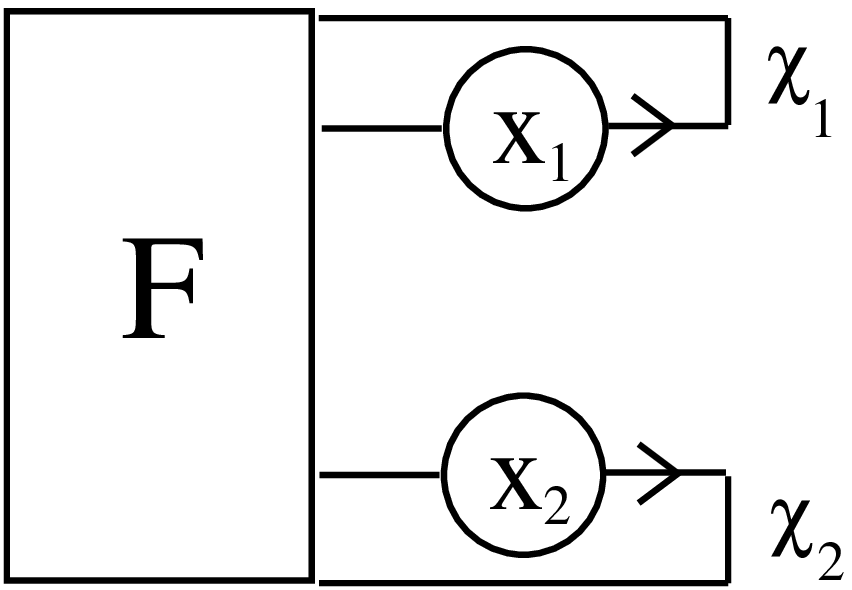}%
}%
%EndExpansion
d\chi_{1}d\chi_{2},
\]
where the box $F$ represents the tensor $F_{z_{1}z_{2}\chi_{1}\chi_{2}%
}^{\acute{z}_{1}\acute{z}_{2}}$. The action of $g\in SO(4,C)$ on $f$ is%

\begin{equation}
gf(x_{1},x_{2})=\iint\limits_{\chi_{1}\chi_{2}}%
%TCIMACRO{\FRAME{itbphF}{2.2935in}{1.0032in}{0.5518in}{}{\Qlb{3}}%
%{xsimplicity2.eps}{\special{ language "Scientific Word";  type "GRAPHIC";
%width 2.2935in;  height 1.0032in;  depth 0.5518in;  original-width 5.5685in;
%original-height 2.4206in;  cropleft "0.002609";  croptop "1";
%cropright "1.002609";  cropbottom "0";
%filename 'Xsimplicity2.eps';file-properties "XNPEU";}}}%
%BeginExpansion
\raisebox{-0.5518in}{\includegraphics[
trim=0.014528in 0.000000in -0.014528in 0.000000in,
height=1.0032in,
width=2.2935in
]%
{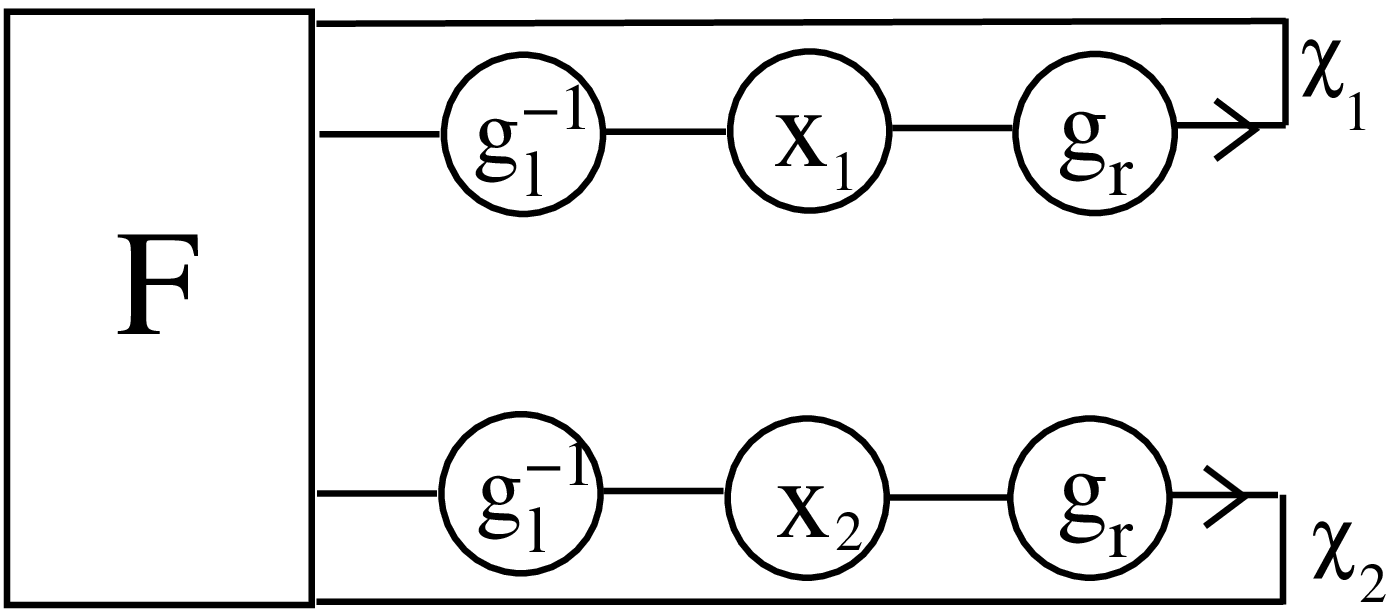}%
}%
%EndExpansion
d\chi_{1}d\chi_{2}. \label{gfresult}%
\end{equation}
Now for any $h\in SL(2,C),$%
\[
T_{a_{1}\chi_{1}}^{b_{1}}(h)T_{a_{2}\chi_{1}}^{b_{2}}(h)=C_{\chi_{1}\chi
_{2}b_{3}}^{b_{1}b_{2}\chi_{3}}\bar{C}_{a_{1}a_{2}\chi_{3}}^{\chi_{1}\chi
_{2}a_{3}}T_{a_{3}\chi_{3}}^{b_{3}}(h),
\]
where $C$'s are the Clebsch-Gordan coefficients of $SL(2,C)$ \cite{IMG},
\cite{NaimarckClebsch}\footnote{I\ derived this equation explicitly in the
appendix of Ref:\cite{MyRigorousSpinFoam}.}. I have assumed all the repeated
indices are either integrated or summed over for the previous and the next two
equations. Using this I can rewrite the $g_{L}$ and $g_{R}$ parts of the
result (\ref{gfresult}) as follows:%
\begin{equation}
T_{a_{1}\chi_{1}}^{z_{1}}(g_{L}^{-1})T_{a_{2}\chi_{2}}^{z_{2}}(g_{L}%
^{-1})=C_{\chi_{1}\chi_{2}z_{3}}^{z_{1}z_{2}\chi_{L}}\bar{C}_{a_{1}a_{2}%
\chi_{L}}^{\chi_{1}\chi_{2}a_{3}}T_{a_{3}\chi_{L}}^{z_{3}}(g_{L}^{-1})
\label{exp1}%
\end{equation}
and%
\begin{equation}
T_{\acute{z}_{1}\chi_{1}}^{b_{1}}(g_{R})T_{\acute{z}_{2}\chi_{2}}^{b_{2}%
}(g_{R})=C_{\chi_{1}\chi_{2}b_{3}}^{b_{1}b_{2}\chi_{R}}\bar{C}_{\acute{z}%
_{1}\acute{z}_{2}\chi_{R}}^{\chi_{1}\chi_{2}\acute{z}_{3}}T_{\acute{z}_{3}%
\chi_{R}}^{b_{3}}(g_{R}). \label{exp2}%
\end{equation}
Now we have
\[
gf(x_{1},x_{2})=\idotsint\limits_{\chi_{1}\chi_{2}\chi_{L}\chi_{R}}%
%TCIMACRO{\FRAME{itbphF}{2.2701in}{1.1588in}{0.5708in}{}{\Qlb{4}}%
%{xsimplicity3.eps}{\special{ language "Scientific Word";  type "GRAPHIC";
%display "USEDEF";  valid_file "F";  width 2.2701in;  height 1.1588in;
%depth 0.5708in;  original-width 7.3388in;  original-height 9.193in;
%cropleft "0";  croptop "1";  cropright "1";  cropbottom "0";
%filename 'Xsimplicity3.eps';file-properties "XNPEU";}}}%
%BeginExpansion
\raisebox{-0.5708in}{\includegraphics[
height=1.1588in,
width=2.2701in
]%
{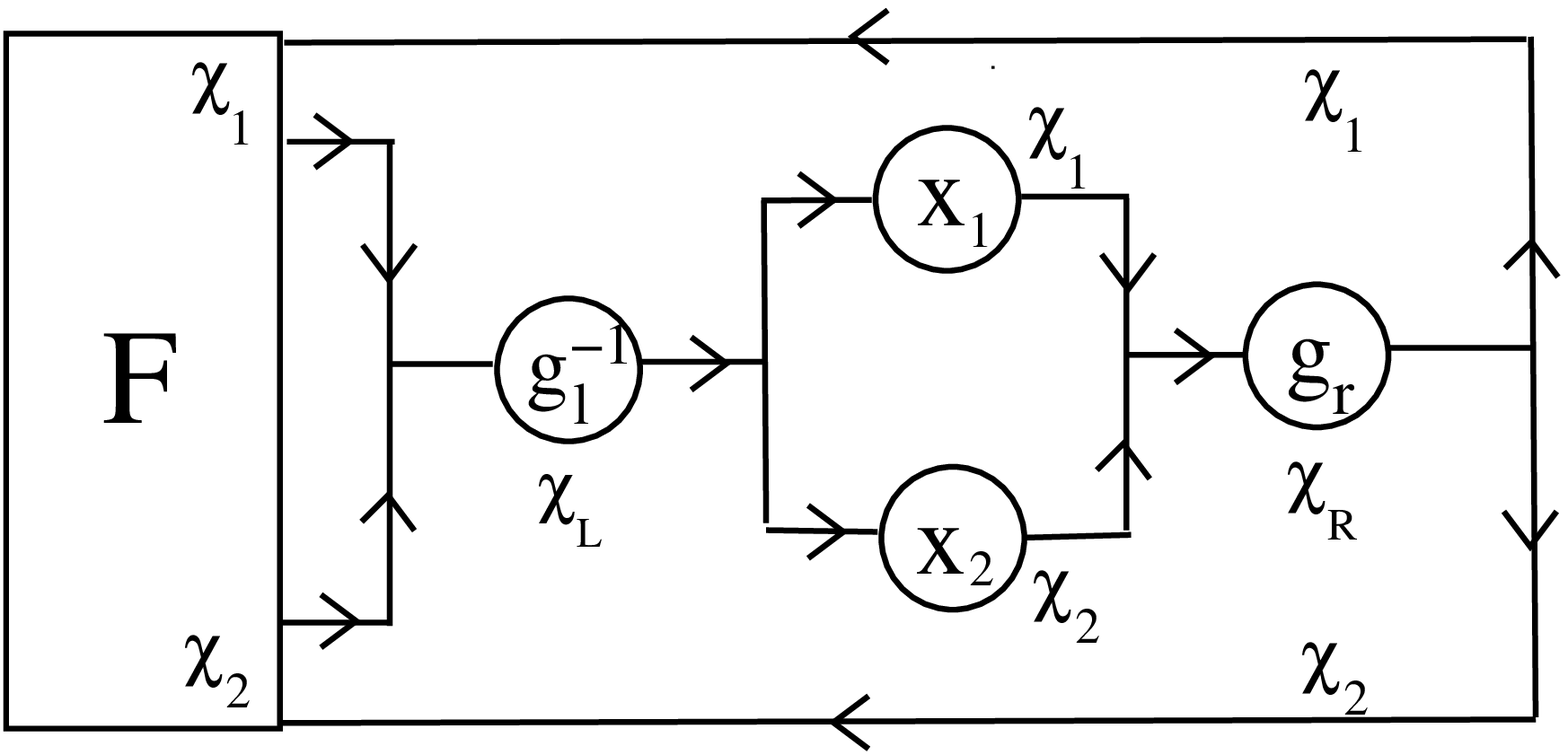}%
}%
%EndExpansion
d\chi_{1}d\chi_{2}.
\]

To satisfy the cross-simplicity constraint the expansion of $gf(x_{1},x_{2})$
must have contribution only from the terms with $\chi_{L}=\pm\chi_{R}$. In the
expansion in equation (\ref{exp1}) and equation (\ref{exp2}) in the right hand
side the terms are defined only up to a sign of $\chi_{L}$ and $\chi_{R}%
$\footnote{Please see appendix A for the explanation.}. Let me remove all the
terms which does not satisfy $\chi_{L}=\pm\chi_{R}$ (say = $\pm\chi$). Also
let me set $g=I.$ Now we can deduce that the functions denoted by $\tilde
{f}(x_{1},x_{2})$ obtained by reducing $f(x_{1},x_{2})$ using the
cross-simplicity constraints must have the expansion \footnote{The factor of 2
has been introduced to include the terms with $\chi_{L}=-\chi_{R}.$},%

\[
f(x_{1},x_{2})=2\iiint\limits_{\chi_{1}\chi_{2}\chi}c_{\chi}%
%TCIMACRO{\FRAME{itbphF}{2.2035in}{1.0006in}{0.5016in}{}{\Qlb{5}}%
%{xsimplicity35.eps}{\special{ language "Scientific Word";  type "GRAPHIC";
%width 2.2035in;  height 1.0006in;  depth 0.5016in;  original-width 7.376in;
%original-height 3.333in;  cropleft "0";  croptop "1";  cropright "1";
%cropbottom "0";  filename 'Xsimplicity35.eps';file-properties "XNPEU";}}}%
%BeginExpansion
\raisebox{-0.5016in}{\includegraphics[
height=1.0006in,
width=2.2035in
]%
{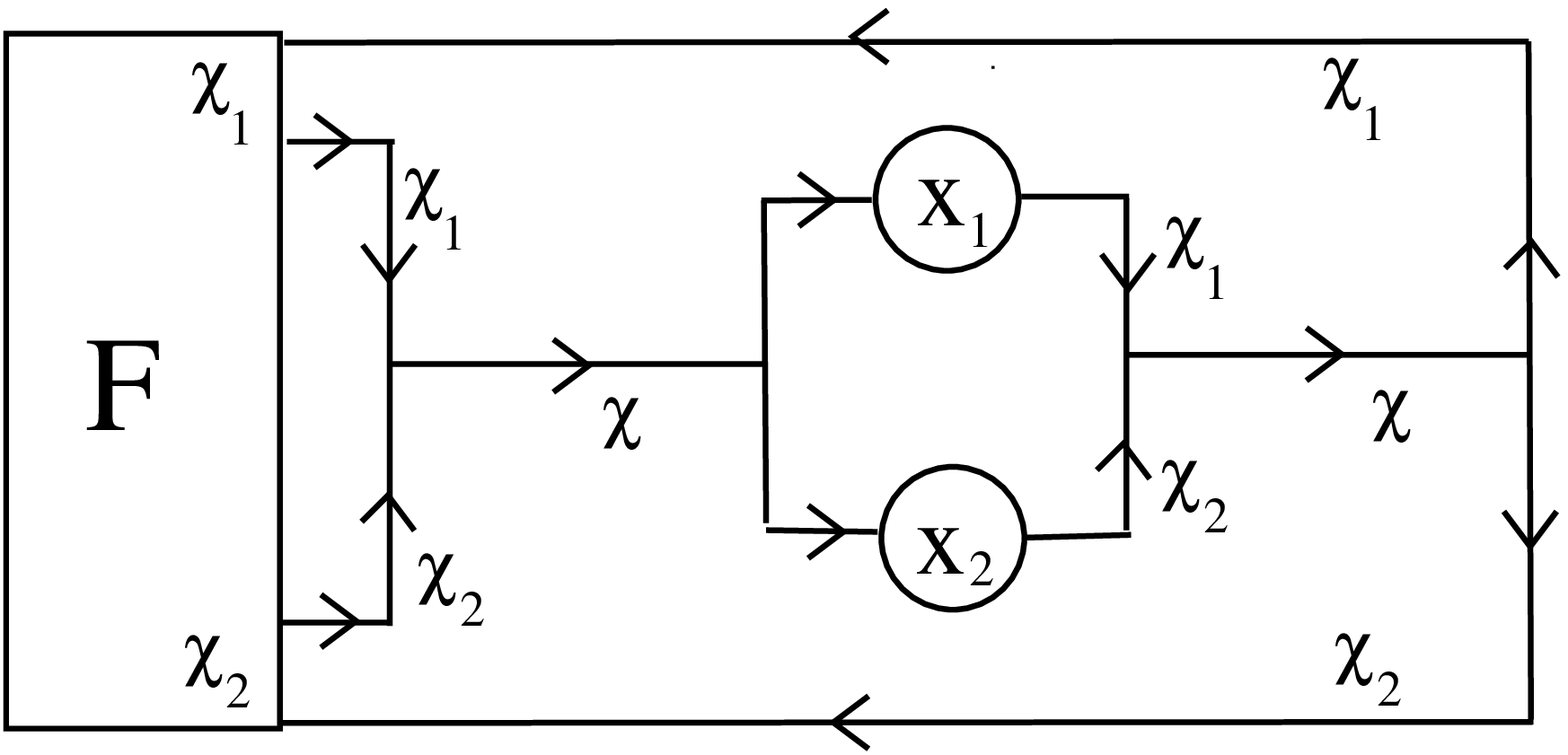}%
}%
%EndExpansion
d\chi_{1}d\chi_{2}d\chi,
\]
where the coefficients $c_{\chi}$ are arbitrary. Now the Clebsch-Gordan
coefficient terms in the expansion can be re-expressed using the following
equation :%
\begin{equation}
C_{\chi_{1}\chi_{2}z_{3}}^{z_{1}z_{2}\chi}\bar{C}_{\acute{z}_{1}\acute{z}%
_{2}\chi}^{\chi_{1}\chi_{2}\acute{z}_{3}}=\frac{8\pi^{4}}{\chi\bar{\chi}}%
\int_{SL(2,C)}T_{\acute{z}_{1}\chi_{1}}^{z_{1}}(h)T_{\acute{z}_{2}\chi_{2}%
}^{z_{2}}(h)\bar{T}_{z_{3}\chi}^{\acute{z}_{3}}(h)dh, \label{clebsmpl}%
\end{equation}
where $h$, $\tilde{h}$ $\in$ $SL(2,C)$ and $dh$ the bi-invariant measure on
$SL(2,C)$. Using this in the middle two Clebsch-Gordan coefficients of
$\tilde{f}(x_{1},x_{2})$ we get
\[
\tilde{f}(x_{1},x_{2})=2\iiint\limits_{\chi_{1}\chi_{2}\chi}\int
_{SL(2,C)}\frac{8\pi^{4}c_{\chi}}{\chi\bar{\chi}}%
%TCIMACRO{\FRAME{itbphF}{1.8983in}{1.0499in}{0.6028in}{}{\Qlb{6}}%
%{xsimplicity4.eps}{\special{ language "Scientific Word";  type "GRAPHIC";
%display "USEDEF";  valid_file "F";  width 1.8983in;  height 1.0499in;
%depth 0.6028in;  original-width 7.3682in;  original-height 9.193in;
%cropleft "0";  croptop "1";  cropright "1";  cropbottom "0";
%filename 'Xsimplicity4.eps';file-properties "XNPEU";}}}%
%BeginExpansion
\raisebox{-0.6028in}{\includegraphics[
height=1.0499in,
width=1.8983in
]%
{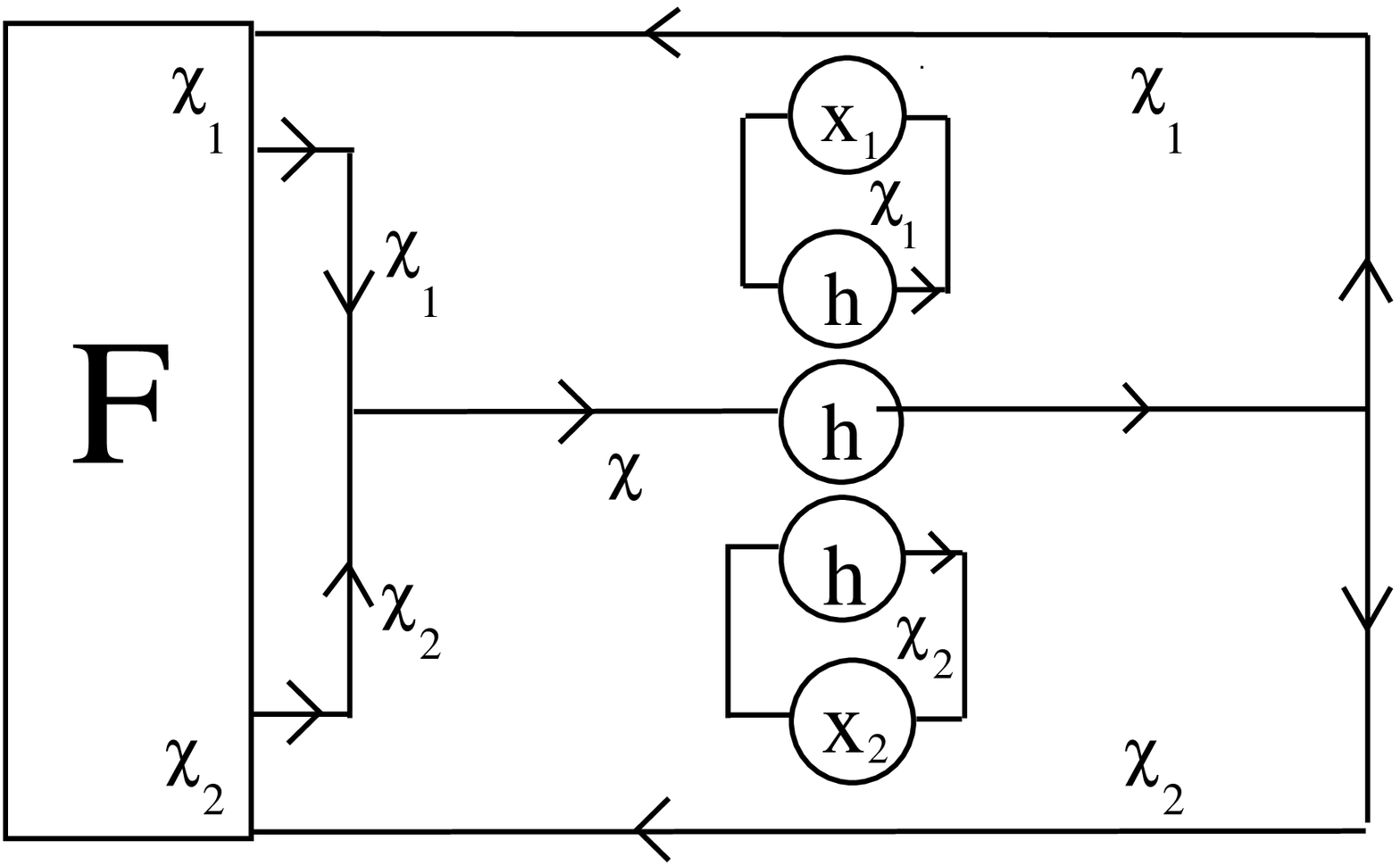}%
}%
%EndExpansion
dhd\chi_{1}d\chi_{2}d\chi.
\]
This result can be rewritten for clarity as%
\[
\tilde{f}(x_{1},x_{2})=2\iiint\limits_{\chi_{1}\chi_{2}\chi}\int
_{SL(2,C)}\frac{8\pi^{4}c_{\chi}}{\chi\bar{\chi}}%
%TCIMACRO{\FRAME{itbphF}{1.3214in}{1.0715in}{0.5016in}{}{\Qlb{7}}%
%{xsimplicity45.eps}{\special{ language "Scientific Word";  type "GRAPHIC";
%width 1.3214in;  height 1.0715in;  depth 0.5016in;  original-width 5.278in;
%original-height 4.273in;  cropleft "0";  croptop "1";  cropright "1";
%cropbottom "0";  filename 'Xsimplicity45.eps';file-properties "XNPEU";}}}%
%BeginExpansion
\raisebox{-0.5016in}{\includegraphics[
height=1.0715in,
width=1.3214in
]%
{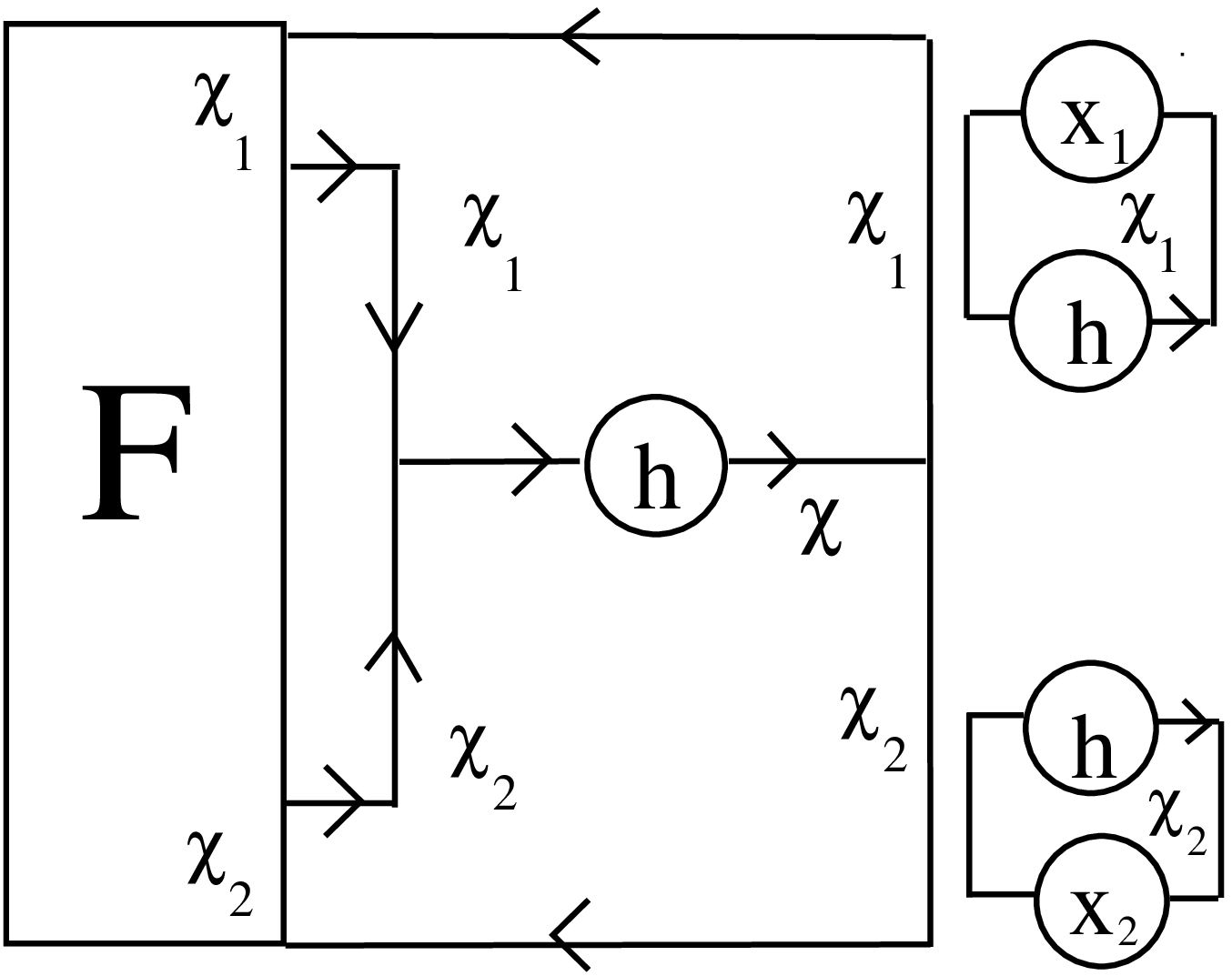}%
}%
%EndExpansion
dhd\chi_{1}d\chi_{2}d\chi.
\]
Once again applying equation (\ref{clebsmpl}) to the remaining two
Clebsch-Gordan coefficients we get,%

\[
\tilde{f}(x_{1},x_{2})=2\iiint\limits_{\chi_{1}\chi_{2}\chi}c_{\chi}%
\iint_{SL(2,C)\times SL(2,C)}\left(  \frac{8\pi^{4}}{\chi\bar{\chi}}\right)
^{2}%
%TCIMACRO{\FRAME{itbphF}{1.203in}{1.0404in}{0.5016in}{}{\Qlb{8}}%
%{xsimplicity5.eps}{\special{ language "Scientific Word";  type "GRAPHIC";
%width 1.203in;  height 1.0404in;  depth 0.5016in;  original-width 4.9571in;
%original-height 4.286in;  cropleft "0";  croptop "1";  cropright "1";
%cropbottom "0";  filename 'Xsimplicity5.eps';file-properties "XNPEU";}}}%
%BeginExpansion
\raisebox{-0.5016in}{\includegraphics[
height=1.0404in,
width=1.203in
]%
{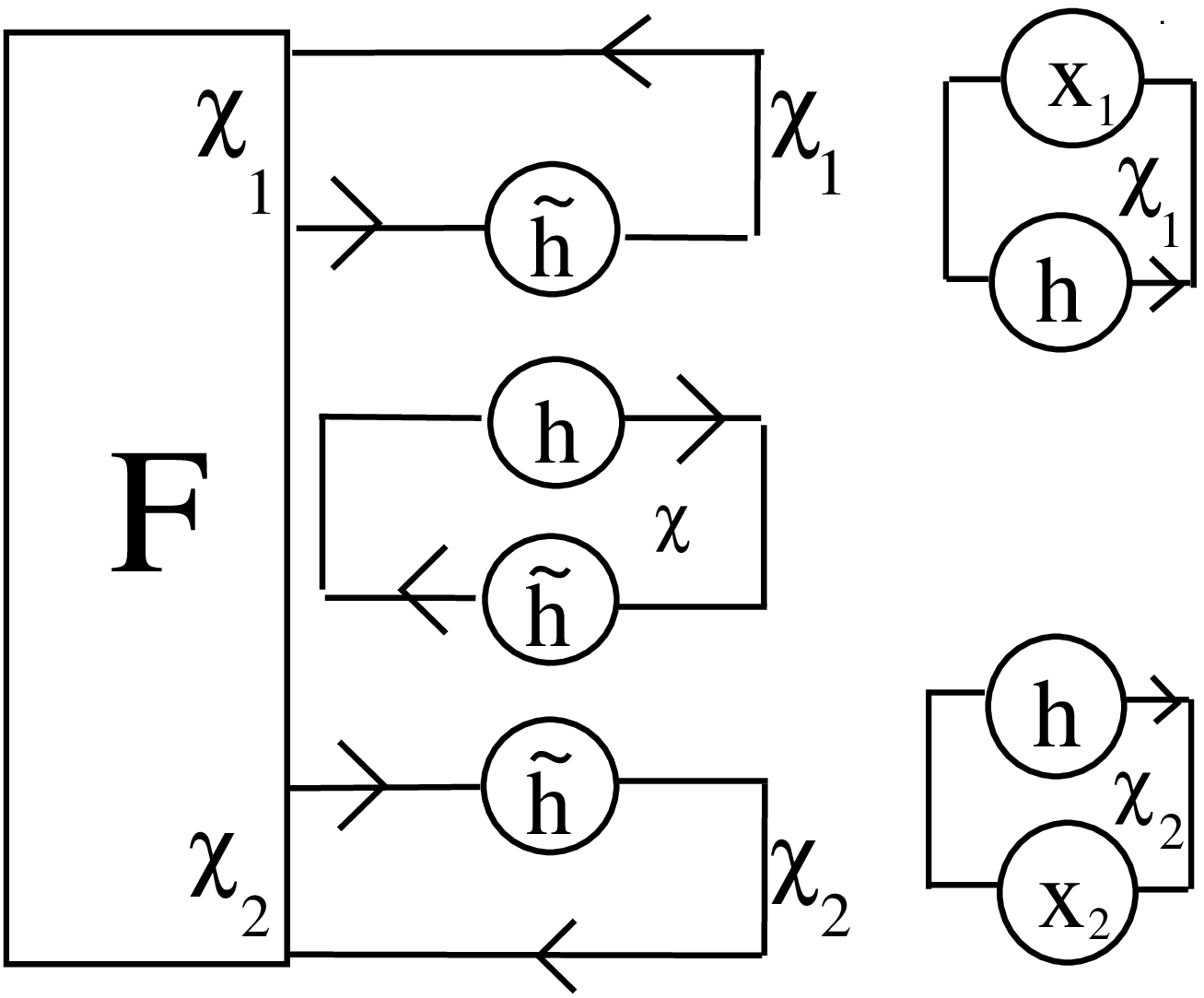}%
}%
%EndExpansion
dhd\tilde{h}d\chi_{1}d\chi_{2}d\chi.
\]

By rewriting the above expression, I deduce that a general function $\tilde
{f}(x_{1},x_{2})$ that satisfies the cross-simplicity constraint must be of
the form,
\begin{subequations}
\label{xsimpfunc}%
\begin{align*}
\tilde{f}(x_{1},x_{2})  &  =\iint\limits_{\chi_{1}\chi_{2}}c_{\chi}%
\int_{SL(2,C)}F_{\chi_{1}\chi_{2}}(h)%
%TCIMACRO{\FRAME{itbphF}{1.5368in}{0.7178in}{0.3684in}{}{\Qlb{9}}%
%{xsimplicity55.eps}{\special{ language "Scientific Word";  type "GRAPHIC";
%display "USEDEF";  valid_file "F";  width 1.5368in;  height 0.7178in;
%depth 0.3684in;  original-width 2.303in;  original-height 1.1727in;
%cropleft "0";  croptop "1";  cropright "1";  cropbottom "0";
%filename 'Xsimplicity55.eps';file-properties "XNPEU";}}}%
%BeginExpansion
\raisebox{-0.3684in}{\includegraphics[
height=0.7178in,
width=1.5368in
]%
{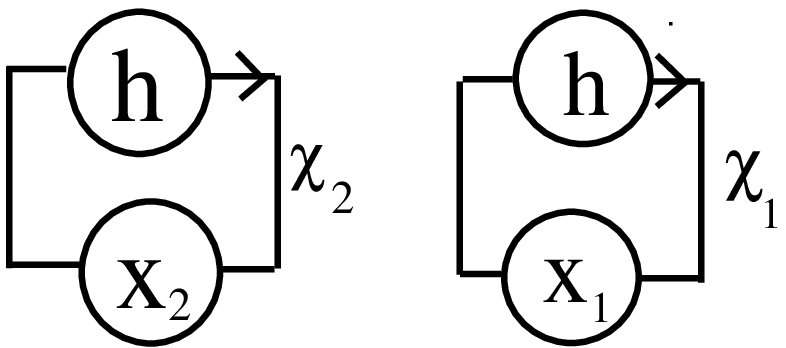}%
}%
%EndExpansion
dhd\chi_{1}d\chi_{2},\\
&  =\iint\limits_{\chi_{1}\chi_{2}}c_{\chi}\int_{SL(2,C)}F_{\chi_{1}\chi_{2}%
}(h)tr(T_{\chi_{1}}(\mathfrak{g}(x_{1})h)tr(T_{\chi_{2}}(\mathfrak{g}%
(x_{2})h)dhd\chi_{1}d\chi_{2},
\end{align*}
where $F_{\chi_{1}\chi_{2}}(h)$ is arbitrary. Then if $\Psi(x_{1},x_{2}%
,x_{3},x_{4})$ is the quantum state of a tetrahedron that satisfies all of the
simplicity constraints and the cross-simplicity constraints, it must be of the
form,
\end{subequations}
\begin{align*}
&  \Psi(x_{1},x_{2},x_{3},x_{4})\\
&  =\int F_{\chi_{1}\chi_{2}\chi_{3}\chi_{4}}(h)tr(T_{\chi_{1}}(\mathfrak{g}%
(x_{1})h)tr(T_{\chi_{2}}(\mathfrak{g}(x_{2})h)\\
&  tr(T_{\chi_{3}}(\mathfrak{g}(x_{3})h)tr(T_{\chi_{4}}(\mathfrak{g}%
(x_{4})h)dh\prod\limits_{i}d\chi_{i}.
\end{align*}
This general form is deduced by requiring that for every pair of variables
with the other two fixed, the function must be the form of the right hand side
of equation (\ref{xsimpfunc}).

\subsubsection{The $SO(4,C)$ Barrett-Crane Intertwiner}

Now the quantization of the fourth Barrett-Crane constraint demands that
$\Psi$ is invariant under the simultaneous complex rotation of its variables.
This is achieved if $F_{\chi_{1}\chi_{2}\chi_{3}\chi_{4}}(h)$ is constant
function of $h$. Therefore the quantum state of a tetrahedron is spanned by%
\begin{equation}
\Psi(x_{1},x_{2},x_{3},x_{4})=\int_{n\in CS^{3}}%
%TCIMACRO{\dprod \limits_{i}}%
%BeginExpansion
{\displaystyle\prod\limits_{i}}
%EndExpansion
T_{\chi_{i}}(\mathfrak{g}(x_{i})\mathfrak{g}(n))dn, \label{BCintertwiner}%
\end{equation}
where the measure $dn$ on $CS^{3}$ is derived from the bi-invariant measure on
$SL(2,C)$. I\ would like to refer to the functions $T_{\chi_{i}}%
(\mathfrak{g}(x_{i})$ as the $T-$\textbf{functions} here after.

\paragraph{Alternative forms}

The quantum state can be diagrammatically represented as follows:%
\[
\Psi(x_{1},x_{2},x_{3},x_{4})=\int%
%TCIMACRO{\FRAME{itbphF}{1.3059in}{1.3059in}{0.6599in}{}{\Qlb{10}%
%}{bcintgroup.eps}{\special{ language "Scientific Word";  type "GRAPHIC";
%width 1.3059in;  height 1.3059in;  depth 0.6599in;  original-width 5.988in;
%original-height 5.988in;  cropleft "0";  croptop "1";  cropright "1";
%cropbottom "0";  filename 'BCIntGroup.eps';file-properties "XNPEU";}}}%
%BeginExpansion
\raisebox{-0.6599in}{\includegraphics[
height=1.3059in,
width=1.3059in
]%
{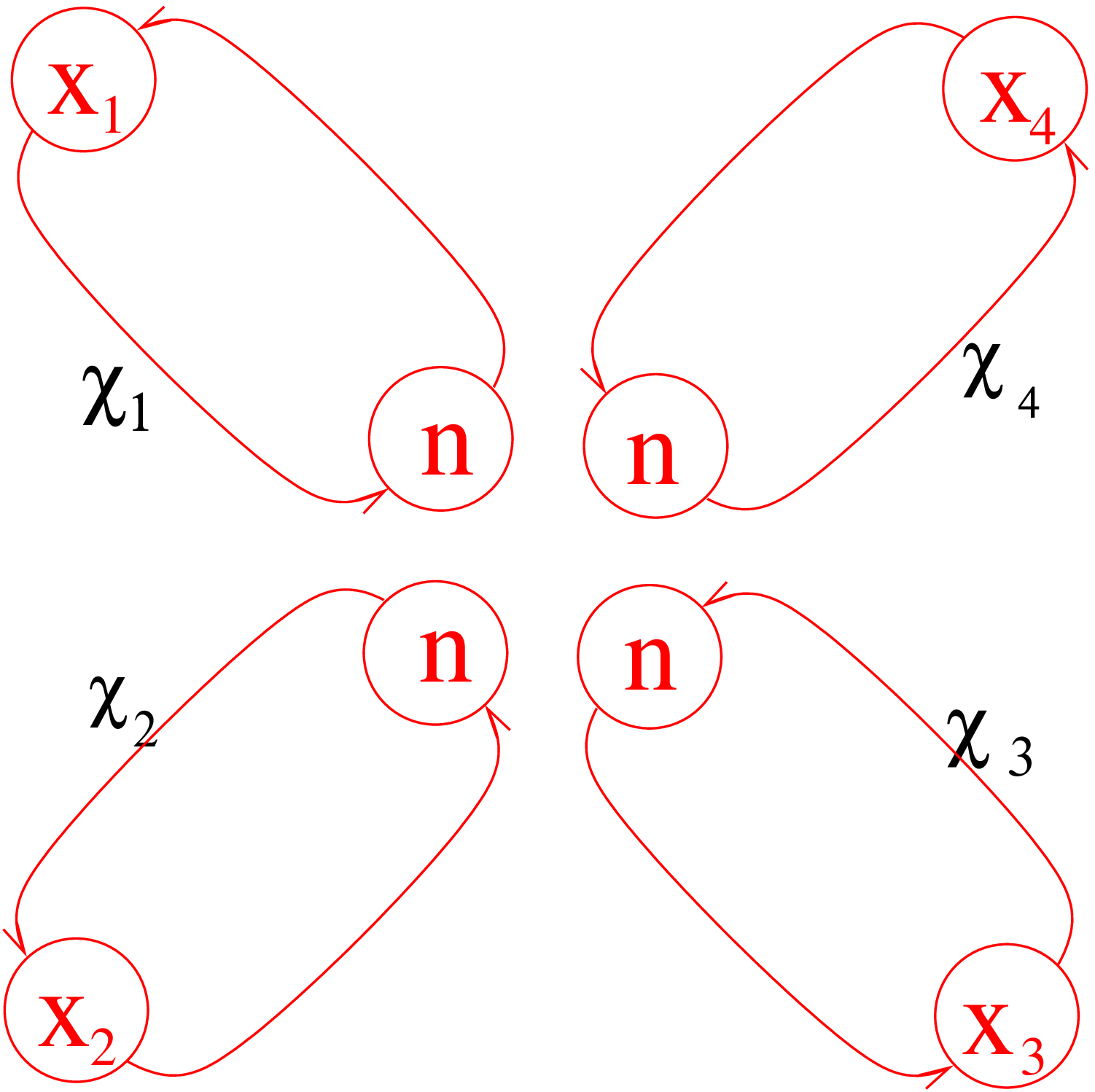}%
}%
%EndExpansion
dn.
\]
A unitary representation $T_{\chi}$ of $SL(2,C)$ can be considered as an
element of $D_{\chi}\otimes D_{\chi}^{\ast}$ where $D_{\chi}^{\ast}$ is the
dual representation of $D_{\chi}$. So using this the Barrett-Crane intertwiner
can be written as an element $\left\vert \Psi\right\rangle \in\bigotimes
\limits_{i}D_{\chi_{i}}\otimes D_{\chi_{i}}^{\ast}$ as follows:%
\[
\left\vert \Psi\right\rangle =\int\limits_{CS^{3}}%
%TCIMACRO{\FRAME{itbphF}{1.3474in}{1.2955in}{0.6529in}{}{\Qlb{11}%
%}{bcintrep.eps}{\special{ language "Scientific Word";  type "GRAPHIC";
%display "USEDEF";  valid_file "F";  width 1.3474in;  height 1.2955in;
%depth 0.6529in;  original-width 6.4688in;  original-height 6.2457in;
%cropleft "0";  croptop "1";  cropright "1";  cropbottom "0";
%filename 'BCIntRep.eps';file-properties "XNPEU";}}}%
%BeginExpansion
\raisebox{-0.6529in}{\includegraphics[
height=1.2955in,
width=1.3474in
]%
{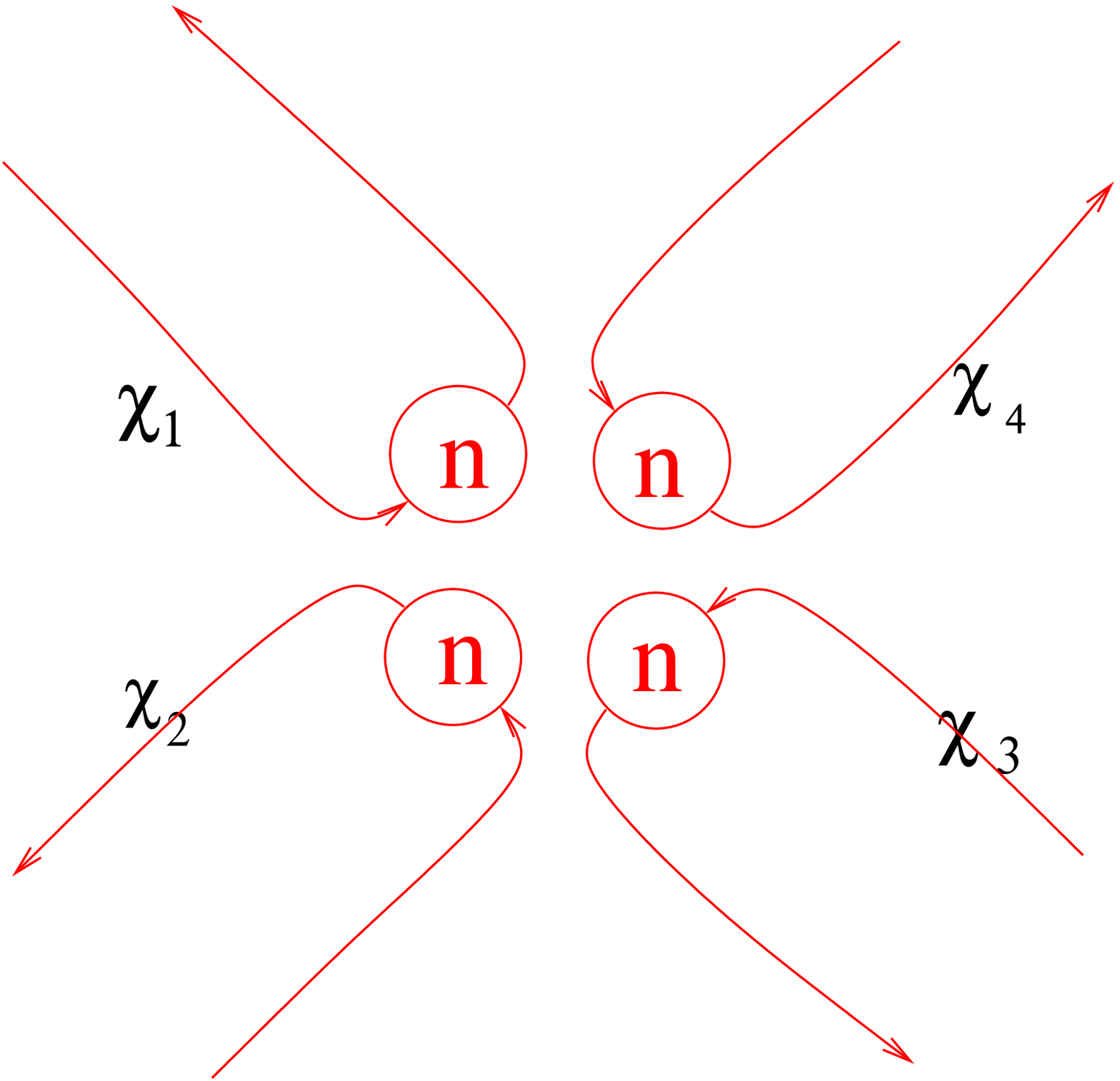}%
}%
%EndExpansion
dn.
\]
Since $SL(2,C)\approx CS^{3},$ using the following graphical identity:%

\[
\int_{SL(2,C)}%
%TCIMACRO{\FRAME{itbphF}{0.9824in}{1.1338in}{0.5708in}{}{\Qlb{12}%
%}{4intertwine.eps}{\special{ language "Scientific Word";  type "GRAPHIC";
%width 0.9824in;  height 1.1338in;  depth 0.5708in;  original-width 2.7319in;
%original-height 3.1531in;  cropleft "0";  croptop "1";  cropright "1";
%cropbottom "0";  filename '4intertwine.eps';file-properties "XNPEU";}}}%
%BeginExpansion
\raisebox{-0.5708in}{\includegraphics[
height=1.1338in,
width=0.9824in
]%
{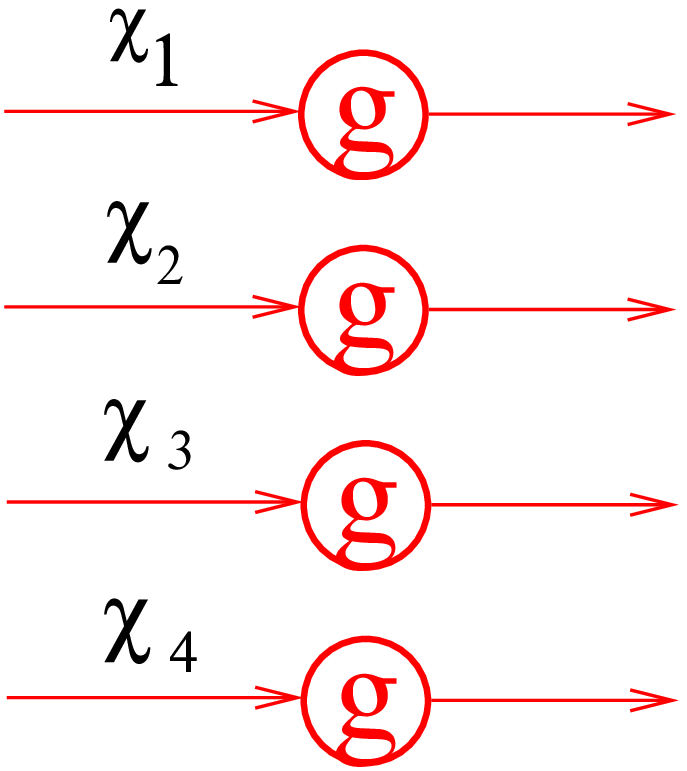}%
}%
%EndExpansion
dg=\int%
%TCIMACRO{\FRAME{itbphF}{2.0877in}{1.1122in}{0.5613in}{}{\Qlb{13}%
%}{edge4int3.eps}{\special{ language "Scientific Word";  type "GRAPHIC";
%width 2.0877in;  height 1.1122in;  depth 0.5613in;  original-width 2.5668in;
%original-height 1.3534in;  cropleft "0";  croptop "1";  cropright "1";
%cropbottom "0";  filename 'edge4int3.eps';file-properties "XNPEU";}}}%
%BeginExpansion
\raisebox{-0.5613in}{\includegraphics[
height=1.1122in,
width=2.0877in
]%
{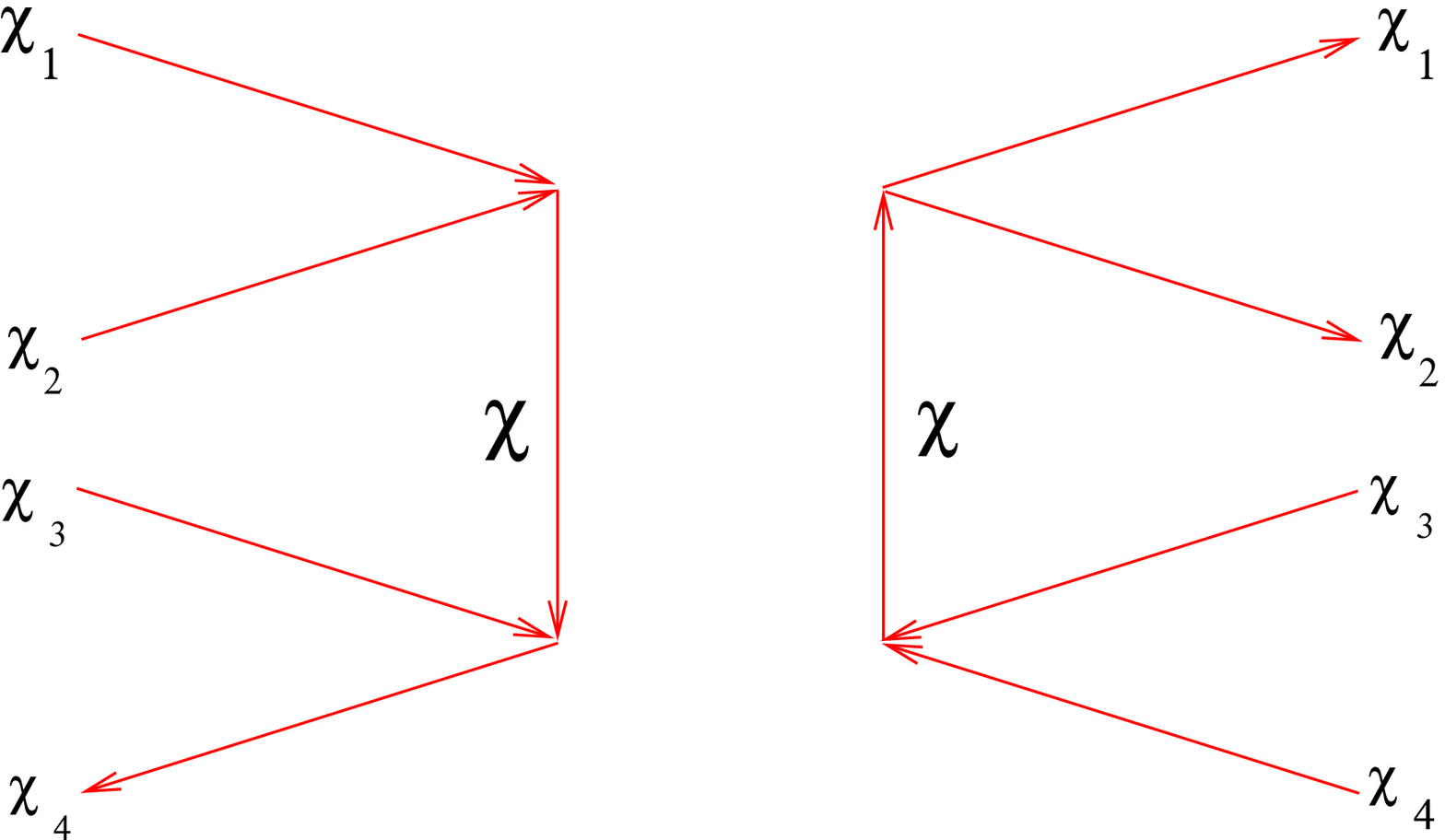}%
}%
%EndExpansion
\frac{8\pi^{4}}{\chi\bar{\chi}}d\chi,
\]
the Barrett-Crane solution can be rewritten as%

\[
\left\vert \Psi\right\rangle =\int%
%TCIMACRO{\FRAME{itbphF}{2.1932in}{0.9617in}{0.4817in}{}{\Qlb{14}}%
%{4xxbc.eps}{\special{ language "Scientific Word";  type "GRAPHIC";
%width 2.1932in;  height 0.9617in;  depth 0.4817in;  original-width 5.2088in;
%original-height 2.2684in;  cropleft "0";  croptop "1";  cropright "1";
%cropbottom "0";  filename '4XXBC.eps';file-properties "XNPEU";}}}%
%BeginExpansion
\raisebox{-0.4817in}{\includegraphics[
height=0.9617in,
width=2.1932in
]%
{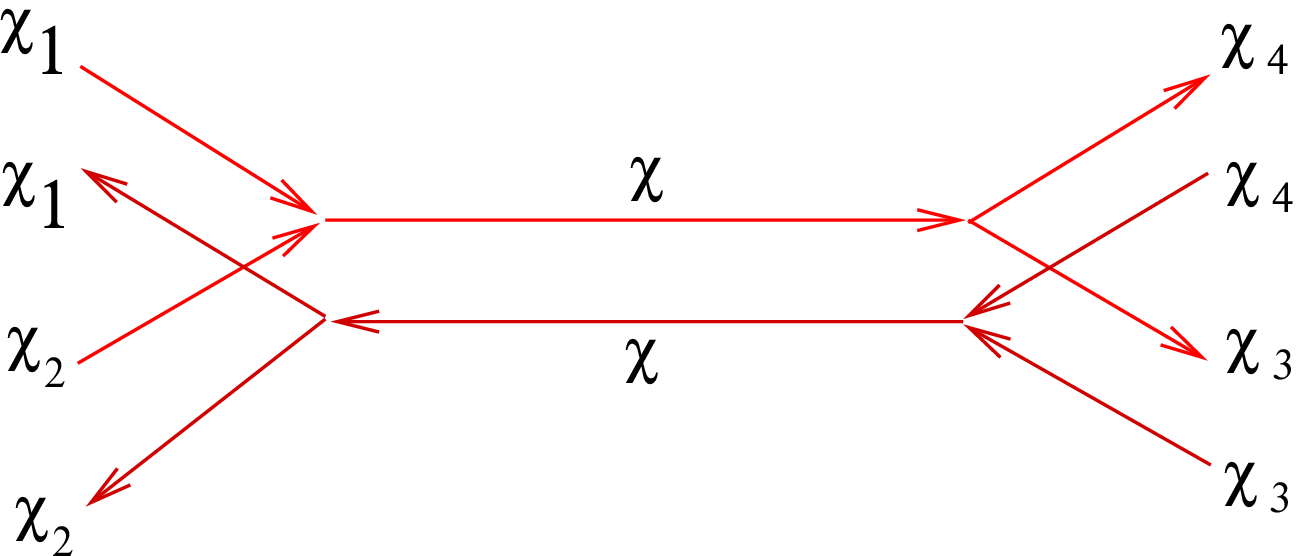}%
}%
%EndExpansion
\frac{8\pi^{4}}{\chi\bar{\chi}}d\chi,
\]
which emerges as an intertwiner in the familiar form in which Barrett and
Crane proposed it for the Riemannian general relativity. It can be clearly
seen that the simple representations for $SO(4,R)$ ($J_{L}=$ $J_{R}$) has been
replaced by the simple representation of $SO(4,C)$ ($\chi_{L}=$ $\pm\chi_{R}$).

\paragraph{Relation to the Riemannian Barrett-Crane model:}

All the analysis done until for the $SO(4,C)\ $Barrett-Crane theory can be
directly applied to the Riemannian$\ $Barrett-Crane theory. The
correspondences between the two models are listed in the following
table\footnote{BC stands for Barrett-Crane. For $\chi_{L}$ and $\chi_{R}$ we
have $n_{L}+n_{R}=even$.
\par
{}}:

\begin{center}%
\begin{tabular}
[c]{lll}%
\textbf{Property} & $SO(4,R)$ BC model & $SO(4,C)$ BC model\\
Gauge group & $SO(4,R)\approx\frac{SL(2,C)\otimes SL(2,C)}{Z_{2}}$ &
$SO(4,C)\approx\frac{SU(2)\otimes SU(2)}{Z_{2}}$\\
Representations & $J_{L},J_{R}$ & $\chi_{L},\chi_{R}$\\
Simple representations & $J_{L}=J_{R}$ & $\chi_{L}=\pm\chi_{R}$\\
Homogenous space & $S^{3}\approx SU(2)$ & $CS^{3}\approx SL(2,C)$%
\end{tabular}

\end{center}

\subsubsection{The Spin Foam Model for the $SO(4,C)$ General Relativity.}

The $SO(4,C)$ Barrett-Crane intertwiner derived in the previous section can be
used to define a $SO(4,C)$ Barrett-Crane spin foam model. The amplitude
$Z_{BC}(s)$ of a four-simplex $s$ is given by the $\{10\chi\}_{SO(4,C)}$
symbol given below:%

\begin{equation}
\{10\chi\}_{SO(4,C)}=%
%TCIMACRO{\FRAME{itbphF}{1.5134in}{1.4131in}{0.7022in}{}{\Qlb{15}}%
%{10xbc.eps}{\special{ language "Scientific Word";  type "GRAPHIC";
%display "USEDEF";  valid_file "F";  width 1.5134in;  height 1.4131in;
%depth 0.7022in;  original-width 5.0254in;  original-height 5.2511in;
%cropleft "0";  croptop "1";  cropright "1";  cropbottom "0";
%filename '10XBC.eps';file-properties "XNPEU";}}}%
%BeginExpansion
\raisebox{-0.7022in}{\includegraphics[
height=1.4131in,
width=1.5134in
]%
{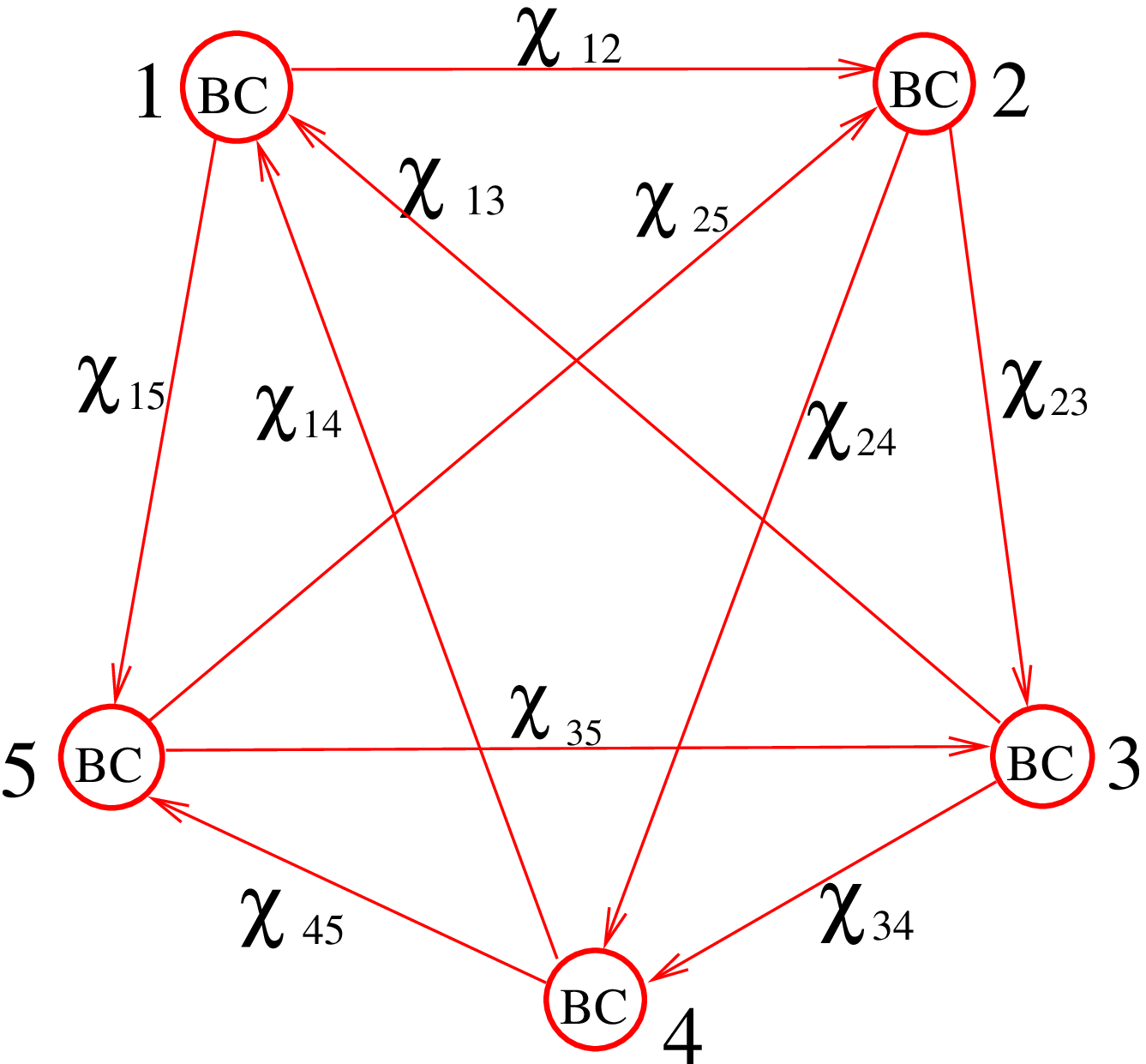}%
}%
%EndExpansion
\text{,} \label{10w}%
\end{equation}
where the circles are the Barrett-Crane intertwiners. The integers represent
the tetrahedra and the pairs of integers represent triangles. The intertwiners
use the four $\chi$'s associated with the links that emerge from it for its
definition in equation (\ref{10w}). In the next subsection, the propagators of
this theory are defined and the $\{10\chi\}$ symbol is expressed in terms of
the propagators in the subsubsection that follows it.

The $SO(4,C)$ Barrett-Crane partition function of the spin foam associated
with the four dimensional simplicial manifold with a triangulation $\Delta$
is
\begin{equation}
Z(\Delta)=\sum_{\left\{  \chi_{b}\right\}  }\left(  \prod_{b}\frac{d_{\chi
_{b}}^{2}}{64\pi^{8}}\right)  \prod_{s}Z(s), \label{PartitionFunction}%
\end{equation}
where $Z(s)$ is the quantum amplitude associated with the $4$-simplex $s$ and
the $d_{\chi_{b}}$ adopted from the spin foam model of the $BF\ $theory can be
interpreted as the quantum amplitude associated with the bone $b$.

\subsubsection{The Features of the $SO(4,C)$ Spin Foam}

\begin{itemize}
\item Areas: The squares of the areas of the triangles (bones) of the
triangulation are given by $\eta_{IK}\eta_{JL}B^{IJ}B^{KL}$. The eigen values
of the squares of the areas in the $SO(4,C)$ Barrett-Crane model from equation
(\ref{eq.2}) are given by
\begin{align*}
\eta_{IK}\eta_{JL}\hat{B}_{b}^{IJ}\hat{B}_{b}^{KL}  &  =\left(  \chi
^{2}-1\right)  \hat{I}\\
&  =\left(  \frac{n^{2}}{2}-\rho^{2}-1+i\rho n\right)  \hat{I}.
\end{align*}
One can clearly see that the area eigen values are complex. The $SO(4,C)$
Barrett-Crane model relates to the $SO(4,C)$ general relativity. Since in the
$SO(4,C)$ general relativity the bivectors associated with any two dimensional
flat object are complex, it is natural to expect that the areas defined in
such a theory are complex too. This is a generalization of the concept of the
space-like and the time-like areas for the real general relativity models:
Area is imaginary if it is time-like and real if it is space-like.

\item Propagators: Laurent and Freidel have investigated the idea of
expressing simple spin networks as Feynman diagrams \cite{SimpleSpinNetworks}.
Here we will apply this idea to the $SO(4,C)$ simple spin networks. Let
$\Sigma$ be a triangulated three surface. Let $n_{i}\in CS^{3}$ be a vector
associated with the $i^{th}$ tetrahedron of the $\Sigma$. The propagator of
the $SO(4,C)$ Barrett-Crane model associated with the triangle $ij$ is given
by%
\begin{align*}
G_{\chi_{ij}}(n_{i},n_{j})  &  =Tr(T_{\chi_{ij}}(g(n_{i}))T_{\chi_{ij}}^{\dag
}(g(n_{j})))\\
&  =Tr(T_{\chi_{ij}}(g(n_{i})g^{-1}(n_{j}))),
\end{align*}
where $\chi_{ij}$ is a representation associated with the triangle common to
the $i^{th}$ and the $j^{th}$ tetrahedron of $\Sigma$. If $X$ and $Y$ belong
to $CS^{3}$ then
\[
tr\left(  \mathfrak{g}(X)\mathfrak{g}(Y)^{-1}\right)  =2X.Y,
\]
where $X.Y$ is the Euclidean dot product and $tr$ is the matrix trace. If
$\lambda=e^{t}$ and $\frac{1}{\lambda}$ are the eigen values of $g(X)g(Y)^{-1}%
$ then,%
\begin{align*}
\lambda+\lambda^{-1}  &  =2X.Y\\
X.Y  &  =\cosh(t).
\end{align*}
From the expression for the trace of the $SL(2,C)$ unitary representations,
(appendix A, \cite{IMG}) I have the propagator for the $SO(4,C)$ Barrett-Crane
model calculated as%
\[
G_{\chi_{ij}}(n_{i},n_{j})=\dfrac{\cos(\rho_{ij}\eta_{ij}+n_{ij}\theta_{ij}%
)}{\left\vert \sinh(\eta_{ij}+i\theta_{ij})\right\vert ^{2}},
\]
where $\eta_{ij}+i\theta_{ij}$ is defined by $n_{i}.n_{j}=\cosh(\eta
_{ij}+i\theta_{ij})$. Two important properties of the propagators are listed below.

\begin{enumerate}
\item Using the expansion for the delta on $SL(2,C)$ I have
\begin{align*}
\delta_{CS^{3}}(X,Y)  &  =\delta_{SL(2,C)}(g(X)g^{-1}(Y))\\
&  =\frac{1}{8\pi^{4}}\int\bar{\chi}\chi Tr(T_{\chi}(g(X)g^{-1}(Y))d\chi,
\end{align*}
where the suffix on the deltas indicate the space in which it is defined.
Therefore%
\[
\int\bar{\chi}\chi G_{\chi}(X,Y))=8\pi^{4}\delta_{CS^{3}}(X,Y).
\]

\item Consider the orthonormality property of the principal unitary
representations of $SL(2,C)$ given by%
\begin{align*}
&  \int_{CS^{3}}T_{\acute{z}_{1}\chi_{1}}^{z_{1}}(g(X))T_{\acute{z}_{2}%
\chi_{2}}^{\dag z_{2}}(g(X))dX\\
&  =\frac{8\pi^{4}}{\chi_{1}\bar{\chi}_{1}}\delta(\chi_{1}-\chi_{2}%
)\delta(z_{1}-\acute{z}_{1})\delta(z_{2}-\acute{z}_{2}),
\end{align*}
where the delta on the $\chi$'s is defined up to a sign of them$.$ From this I
have%
\[
\int_{CS^{3}}G_{\chi_{1}}(X,Y)G_{\chi_{2}}(Y,Z)dY=\frac{8\pi^{4}}{\chi_{1}%
\bar{\chi}_{1}}\delta(\chi_{1}-\chi_{2})G_{\chi_{1}}(X,Z).
\]

\end{enumerate}

\item The $\{10\chi\}$ symbol can be defined using the propagators on the
complex three sphere as follows:%
\begin{align*}
Z(s)  &  =\int_{x_{k}\in CS^{3}}%
%TCIMACRO{\dprod \limits_{i<j}}%
%BeginExpansion
{\displaystyle\prod\limits_{i<j}}
%EndExpansion
T_{\chi_{ij}}(\mathfrak{g}(x_{i})\mathfrak{g}(x_{j}))%
%TCIMACRO{\dprod \limits_{k}}%
%BeginExpansion
{\displaystyle\prod\limits_{k}}
%EndExpansion
dx_{k},\\
&  =\int_{\forall x_{k}\in CS^{3}}%
%TCIMACRO{\dprod \limits_{i<j}}%
%BeginExpansion
{\displaystyle\prod\limits_{i<j}}
%EndExpansion
G_{\chi_{ij}}(x_{i}\mathfrak{,}x_{j})%
%TCIMACRO{\dprod \limits_{k}}%
%BeginExpansion
{\displaystyle\prod\limits_{k}}
%EndExpansion
dx_{k},
\end{align*}
where $i$ denotes a tetrahedron of the four-simplex. For each tetrahedron
$k,~$a free variable $x_{k}\in CS^{3}$ is associated. For each triangle $ij$
which is the intersection of the $i$'th and the $j$'th tetrahedron, a
representation of $SL(2,C)$ denoted by $\chi_{ij}$ is associated.

\item Discretization Dependence and Local Excitations: It is well known that
the BF theory is discretization independent and is topological. The$\ $spin
foam for the $SO(4,C)$ general relativity is got by imposing the Barrett-Crane
constraints on the $BF\ $Spin foam. After the imposition of the Barrett-Crane
constraints the theory loses the discretization independence and the
topological nature. This can be seen in many ways.

\begin{itemize}
\item The simplest reason is that the $SO(4,C)\ $Barrett-Crane model
corresponds to the quantization of the discrete $SO(4,C)$ general relativity
which has local degrees of freedom.

\item After the restriction of the representations involved in BF spin foams
to the simple representations and the intertwiners to the Barrett-Crane
intertwiners, various important identities used in the spin foam diagrammatics
and proof of the discretization independence of the BF\ theory spin foams in
Ref:\cite{SpinFoamDiag} are no longer available.

\item The BF partition function is simply gauge invariant measure of the
volume of space of flat connections. Consider the following harmonic expansion
of the delta function which was used in the derivation of the $SO(4,C)$ BF
theory:%
\[
\delta(g)=\frac{1}{8\pi^{4}}\int d_{\omega}tr(T_{\omega}(g))d\omega.
\]
Imposition of the Barrett-Crane constraints on the BF theory spin foam,
suppresses the terms corresponding to the non-simple representations. If only
the simple representations are allowed in the right hand side, it is no longer
peaked at the identity. This means that the partition function for
the\ $SO(4,C)$\ Barrett-Crane model involves contributions only from the
non-flat connections which has local information.

\item In the asymptotic limit study of the $SO(4,C)$ spin foams in section
four the discrete version of the $SO(4,C)$ general relativity (Regge calculus)
is obtained. The Regge calculus action is clearly discretization dependent and non-topological.
\end{itemize}
\end{itemize}

\section{The Asymptotic Limit of the $SO(4,C)$ Barrett-Crane models.}

The asymptotic limit of the real Barrett-Crane models has been studied before
\cite{JWBCS}, \cite{JWBRW}, \cite{PonzanoReggeModel}, \cite{BaezEtalAsym} to a
certain degree. Here I will discuss the asymptotic limit of the $SO(4,C)$
Barrett-Crane model. For the first time I show here that we can extract
bivectors which satisfy the essential Barrett-Crane constraints from the
asymptotic limit. Consider the amplitude of a four-simplex given by Eq.
(\ref{10w}) with a real scale parameter $\lambda$,%
\begin{align*}
Z_{\lambda}  &  =\int_{n_{k}\in CS^{3}}%
%TCIMACRO{\dprod \limits_{i<j}}%
%BeginExpansion
{\displaystyle\prod\limits_{i<j}}
%EndExpansion
G_{\lambda\chi_{ij}}(n_{i}\mathfrak{,}n_{j})%
%TCIMACRO{\dprod \limits_{k}}%
%BeginExpansion
{\displaystyle\prod\limits_{k}}
%EndExpansion
dn_{k},\\
&  =\int_{n_{k}\in CS^{3}}%
%TCIMACRO{\dprod \limits_{i<j}}%
%BeginExpansion
{\displaystyle\prod\limits_{i<j}}
%EndExpansion
\dfrac{\cos(\lambda\rho_{ij}\eta_{ij}+\lambda n_{ij}\theta_{ij})}{\left\vert
\sinh(\lambda\eta_{ij}+i\lambda\theta_{ij})\right\vert ^{2}}%
%TCIMACRO{\dprod \limits_{k}}%
%BeginExpansion
{\displaystyle\prod\limits_{k}}
%EndExpansion
dx_{k},\\
&  =\int_{n_{k}\in CS^{3}}%
%TCIMACRO{\dprod \limits_{i<j}}%
%BeginExpansion
{\displaystyle\prod\limits_{i<j}}
%EndExpansion
\sum\limits_{\varepsilon_{ij}=\pm1}\dfrac{\exp(i\varepsilon_{ij}\lambda
(\rho_{ij}\eta_{ij}+n_{ij}\theta_{ij}))}{2\left\vert \sinh(\lambda\eta
_{ij}+i\lambda\theta_{ij})\right\vert ^{2}}%
%TCIMACRO{\dprod \limits_{k}}%
%BeginExpansion
{\displaystyle\prod\limits_{k}}
%EndExpansion
dx_{k},
\end{align*}
where the $\eta_{ij}+i\theta_{ij}$ is defined by $n_{i}.n_{j}=\cosh(\eta
_{ij}+i\theta_{ij})$. Here the $\zeta_{ij}=$ $\eta_{ij}+i\theta_{ij}$ is the
complex angle between $n_{i}$ and $n_{j}$. The asymptotic limit of
$Z_{\lambda}(s)\ $under $\lambda\longrightarrow\infty$ is controlled by%

\begin{align*}
&  S(\{n_{i},\bar{n}_{i}\},\{\chi_{ij},\bar{\chi}_{ij}\})\\
&  =\sum_{i<j}\varepsilon_{ij}(\rho_{ij}\eta_{ij}+n_{ij}\theta_{ij}%
)+\operatorname{Re}\left(  \sum_{i}q_{i}(n_{i}.n_{i}-1)\right) \\
&  =\operatorname{Re}\left(  \sum_{i<j}\varepsilon_{ij}\bar{\chi}_{ij}%
\zeta_{ij}+\sum_{i}q_{i}(n_{i}.n_{i}-1)\right)  ,
\end{align*}
where the $q_{i}$ are the Lagrange multipliers to impose $n_{i}.n_{i}%
=1,\forall i$. My goal now is to find stationary points for this action. The
stationary points are determined by
\begin{subequations}
\begin{equation}
\sum_{~~~~~i\neq j}\varepsilon_{ij}\bar{\chi}_{ij}\frac{\partial\zeta_{ij}%
}{\partial n_{i}}+q_{j}n_{j}=0,\forall j, \label{ext}%
\end{equation}
and $n_{j}.n_{j}=1,\forall j$ where the $j$ is a constant in the summation.%

\end{subequations}
\begin{equation}
\frac{\partial\zeta_{ij}}{\partial n_{i}}=\frac{n_{j}}{\sinh(\zeta_{ij})}.
\label{eq.dif.as}%
\end{equation}

Using equation (\ref{eq.dif.as}) in equation :(\ref{ext}) and taking the wedge
product of the equation with $n_{j}$ we have,
\[
\sum_{~~~~~i\neq j}\varepsilon_{ij}\bar{\chi}_{ij}\frac{n_{i}\wedge n_{j}%
}{\sinh(\zeta_{ij})}=0,\forall j.
\]

If
\[
\bar{E}_{ij}=i\varepsilon_{ij}\bar{\chi}_{ij}\frac{n_{i}\wedge n_{j}}%
{\sinh(\zeta_{ij})},
\]
then the last equation can be simplified to%

\begin{equation}
\sum_{~~~~~i\neq j}E_{ij}=0,\forall j. \label{sumzero}%
\end{equation}

We now consider the properties of $E_{ij}:$

\begin{itemize}
\item Each $i$ represents a tetrahedron. There are ten $E_{ij}$'s, each one of
them is associated with one triangle of the four-simplex.

\item The square of $E_{ij}$:%
\begin{align*}
\bar{E}_{ij}\cdot\bar{E}_{ij}  &  =\frac{-\bar{\chi}_{ij}^{2}}{\sinh^{2}%
(\zeta_{ij})}(n_{j}^{2}n_{i}^{2}-\left(  n_{i}\cdot n_{j}\right)  ^{2})\\
&  =\frac{-\bar{\chi}_{ij}^{2}}{\sinh^{2}(\zeta_{ij})}(1-\left(  \cosh
(\zeta_{ij}\right)  ^{2})\\
&  =\bar{\chi}_{ij}^{2}.
\end{align*}

\item The wedge product of any two $E_{ij}$ is zero if they are equal to each
other or if their corresponding triangles belong to the same tetrahedron.

\item Sum of all the $E_{ij}$ belonging to the same tetrahedron are zero
according to equation (\ref{sumzero}).
\end{itemize}

It is clear that these properties contain the first four Barrett-Crane
constraints. So we have successfully extracted the bivectors corresponding to
the triangles of a general flat four-simplex in $SO(4,C)$ general relativity
and the $n_{i}$ are the normal vectors of the tetrahedra. The $\chi_{ij}$ are
the complex areas of the triangle as one would expect. Since we did not impose
any non-degeneracy conditions, it is not guaranteed that the tetrahedra or the
four-simplex have non-zero volumes.

The asymptotic limit of the partition function of the entire simplicial
manifold with triangulation $\Delta$ is%
\[
S(\Delta,\{n_{is}\in CS^{3},\chi_{ij},\bar{\chi}_{ij},\varepsilon
_{ijs}\})=\operatorname{Re}\sum_{i<j,s}\varepsilon_{ijs}\bar{\chi}_{ij}%
\zeta_{ijs},
\]
where I\ have assumed variable $s$ represents the four simplices of $\Delta$
and $i,$ $j$ represents the tetrahedra. The $\varepsilon_{ijs}$ can be
interpreted as the orientation of the triangles. Each triangle has a
corresponding $\chi_{ij}$. The $n_{is}$ denote the unit complex vector
associated with the side of the tetrahedron $i$ facing the inside of a simplex
$s$. Now there is one bivector $E_{sij}$ associated with each side facing
inside of a simplex $s$ of a triangle $ij$ defined by%

\[
\bar{E}_{ijs}=i\varepsilon_{ijs}\bar{\chi}_{ij}\frac{n_{i}\wedge n_{js}}%
{\sinh(\zeta_{ijs})}.
\]
If the $n_{is}$ are chosen such that they satisfy stationary conditions%

\[
\sum_{~~~~~i\neq j}E_{ijs}=0,\forall j,s,
\]
and if
\[
\theta_{ij}=\left(  \sum_{s}\varepsilon_{ijs}\zeta_{ijs}\right)  ,
\]

then%

\begin{align*}
S(\Delta,\{\chi_{ij},\bar{\chi}_{ij},\varepsilon_{ijs}\}) &
=\operatorname{Re}\sum_{i<j,s}\varepsilon_{ijs}\bar{\chi}_{ij}\zeta_{ijs},\\
&  =\operatorname{Re}\sum_{i<j}\bar{\chi}_{ij}\theta_{ij}%
\end{align*}
can be considered to describe the Regge calculus for the $SO(4,C)$ general
relativity. The angle $\theta_{ij}$ are the deficit angles associated with the
triangles and the $n_{is}$ are the complex vector normals associated with the
tetrahedra. From the analysis that has been done in this section, it is easy
see that the $SO(4,C)$ Regge calculus contains the Regge calculus theories for
all the signatures. The Regge calculus for each signature can be obtained by
restricting the $n_{is}$ and the $\chi_{ij}$ to the corresponding homogenous
space and representations \cite{RealitySpinFoam}. Also by the properly
restricting the $n_{is}$ and the $\chi_{ij}$ we can derive the Regge calculus
corresponding to the mixed Lorentzian and multi-signature Barrett-Crane models
described in the previous subsections. The details of the relation between the
$SO(4,C)$ Regge Calculus and the real Regge Calculus for different signature
will be studied elsewhere.

\appendix

\section{Unitary Representations of SL(2,$\boldsymbol{C}$)}

The Representation theory of $SL(2,\boldsymbol{C})$ was developed by Gelfand
and Naimarck \cite{IMG}. Representation theory of $SL(2,C)$ can be developed
using functions on $C^{2}$ which are homogenous in their
arguments\footnote{These functions need not be holomorphic but infinitely
differentiable may be except at the origin $(0,0)$.}. The space of functions
$D_{\chi}$ is defined as functions $f(z_{1},z_{2})$ on $C^{2}$ whose
homogeneity is described by%
\[
f(az_{1},az_{2})=a^{\chi_{1}-1}a^{\chi_{2}-1}f(z_{1},z_{2}),
\]
for all $a\neq0,$ where $\chi$ is a pair $(\chi_{1},\chi_{2})$. The linear
action of $SL(2,C)$ on $C^{2}$ defines a representation of $SL(2,C)$ denoted
by $T_{\chi}$. Because of the homogeneity of functions of $D_{\chi},$ the
representations $T_{\chi}$ can be defined by its action on the functions
$\phi(z)$ of one complex variable related to $f(z_{1},z_{2})\in$ $D_{\chi}$ by%
\[
\phi(z)=f(z,1).
\]
There are two qualitatively different unitary representations of $SL(2,C)$:
the principal series and the supplementary series, of which only the first one
is relevant to quantum general relativity. The principal unitary irreducible
representations of $SL(2,\boldsymbol{C})$ are the infinite dimensional. For
these $\chi_{1}=-\bar{\chi}_{2}=\frac{n+i\rho}{2},$ where $n$ is an integer
and $\rho$ is a real number. In this article I\ would like to label the
representations by a single complex number $\chi=\frac{n}{2}+i\frac{\rho}{2}$,
wherever necessary. The $T_{\chi}$ representations are equivalent to
$T_{-\chi}$ representations \cite{IMG}.

Let $g$ be an element of $SL(2,\boldsymbol{C})$ given by%
\[
g=\left[
\begin{array}
[c]{cc}%
\alpha & \beta\\
\gamma & \delta
\end{array}
\right]  ,
\]
where $\alpha$,$\beta$,$\gamma$ and $\delta$ are complex numbers such that
$\alpha\delta-\beta\delta=1$. Then the $D\chi$ representations are described
by the action of a unitary operator $T_{\chi}(g)$ on the square integrable
functions $\phi(z)$ of a complex variable $z$ as given below:%
\begin{equation}
T_{\chi}(g)\phi(z)=(\beta z_{1}+\delta)^{\chi-1}(\bar{\beta}\bar{z}_{1}%
+\bar{\delta})^{-\bar{\chi}-1}\phi(\frac{\alpha z+\gamma}{\beta z+\delta}).
\label{rep}%
\end{equation}
This action on $\phi(z)$ is unitary under the inner product defined by%
\[
\left(  \phi(z),\eta(z)\right)  =\int\bar{\phi}(z)\eta(z)d^{2}z,
\]
where $d^{2}z=\frac{i}{2}dz\wedge d\bar{z}$ and I\ would like to adopt this
convention everywhere. Completing $D_{\chi}$ with the norm defined by the
inner product makes it into a Hilbert space $H_{\chi}$.

Equation (\ref{rep}) can also be written in kernel form
\cite{RovPerGFTLorentz},%
\[
T_{\chi}(g)\phi(z_{1})=\int T_{\chi}(g)(z_{1},z_{2})\phi(z_{2})d^{2}z_{2},
\]
Here $T_{\chi}(g)(z_{1},z_{2})$ is defined as%
\begin{equation}
T_{\chi}(g)(z_{1},z_{2})=(\beta z_{1}+\delta)^{\chi-1}(\bar{\beta}\bar{z}%
_{1}+\bar{\delta})^{-\bar{\chi}-1}\delta(z_{2}-g(z_{1})), \label{eq.rep}%
\end{equation}
where $g(z_{1})=\frac{\alpha z_{1}+\gamma}{\beta z_{1}+\delta}$. The Kernel
$T_{\chi}(g)(z_{1},z_{2})$ is the analog of the matrix representation of the
finite dimensional unitary representations of compact groups. An infinitesimal
group element, $a$, of $SL(2,\boldsymbol{C})$ can be parameterized by six real
numbers $\varepsilon_{k}$ and $\eta_{k}$ as follows \cite{Ruhl}:%
\[
a\approx I+\frac{i}{2}\sum_{k=1}^{3}(\varepsilon_{k}\sigma_{k}+\eta_{k}%
i\sigma_{k}),
\]
where the $\sigma_{k}$ are the Pauli matrices. The corresponding six
generators of the $\chi$ representations are the $H_{k}$ and the $F_{k}$. The
$H_{k}$ correspond to rotations and the $F_{k}$ correspond to boosts. The
bi-invariant measure on $SL(2,C)$ is given by
\[
dg=\left(  \frac{i}{2}\right)  ^{3}\frac{d^{2}\beta d^{2}\gamma d^{2}\delta
}{\left\vert \delta\right\vert ^{2}}=\left(  \frac{i}{2}\right)  ^{3}%
\frac{d^{2}\alpha d^{2}\beta d^{2}\gamma}{\left\vert \alpha\right\vert ^{2}}.
\]
This measure is also invariant under inversion in $SL(2,\boldsymbol{C})$. The
Casimir operators for $SL(2,C$ $)$ are given by%
\[
\hat{C}=\det\left[
\begin{array}
[c]{cc}%
\hat{X}_{3} & \hat{X}_{1}-i\hat{X}_{2}\\
\hat{X}_{1}+i\hat{X}_{2} & -\hat{X}_{3}%
\end{array}
\right]
\]
and its complex conjugate $\bar{C}$ where $X_{i}=F_{i}+iH_{i}.$ The action of
$C$ ($\bar{C}$) on the elements of $D_{\chi}$ reduces to multiplication by
$\chi_{1}^{2}-1$ ($\chi_{2}^{2}-1$).The real and imaginary parts of $C$ are
another way of writing the Casimirs. On $D_{\chi}$ they reduce to the
following%
\begin{align*}
\operatorname{Re}(\hat{C})  &  =\left(  -\rho^{2}+\frac{n}{4}^{2}-1\right)
\hat{I},\\
\operatorname{Im}(\hat{C})  &  =\rho n\hat{I}.
\end{align*}

The Fourier transform theory on $SL(2,\boldsymbol{C})$ was developed in
Ref:\cite{IMG}. If $f(g)$ is a square integrable function on the group, it has
a group Fourier transform defined by%
\begin{equation}
F(\chi)=\int f(g)T_{\chi}(g)dg, \label{Four}%
\end{equation}
where is $F(\chi)$ is linear operator defined by the kernel $K_{\chi}%
(z_{1},z_{2})$ as follows:%
\[
F(\chi)\phi(z)=\int K_{\chi}(z,\acute{z})\phi(\acute{z})d^{2}\acute{z}.
\]
The associated inverse Fourier transform is%
\begin{equation}
f(g)=\frac{1}{8\pi^{4}}\int Tr(F(\chi)T_{\chi}(g^{-1}))\chi\bar{\chi}d\chi,
\label{invFour}%
\end{equation}
where the $\int d\chi$ indicates the integration over $\rho$ and the summation
over $n.$ From the expressions for the Fourier transforms, I can derive the
orthonormality property of the $T_{\chi}$ representations,%
\[
\int_{SL(2,C)}T_{\acute{z}_{1}\chi_{1}}^{z_{1}}(g)T_{\acute{z}_{2}\chi_{2}%
}^{\dag z_{2}}(g)dg=\frac{8\pi^{4}}{\chi_{1}\bar{\chi}_{1}}\delta(\chi
_{1}-\chi_{2})\delta(z_{1}-\acute{z}_{1})\delta(z_{2}-\acute{z}_{2}),
\]
where $T_{\chi}^{\dagger}$ is the Hermitian conjugate of $T_{\chi}$.

The Fourier analysis on $SL(2,C)$ can be used to study the Fourier analysis on
the complex three sphere $CS^{3}$. If $x=(a,b,c,d)\in$ $CS^{3}$ then the
isomorphism $\mathfrak{g}:CS^{3}\longrightarrow SL(2,C)\ $can be defined by
the following:%
\[
\mathfrak{g}(x)=\left[
\begin{array}
[c]{cc}%
a+ib & c+id\\
-c+id & a-ib
\end{array}
\right]  .
\]
Then, the Fourier expansion of $f(x)$ $\in L^{2}(CS^{3})$ is given by%
\[
f(x)=\frac{1}{8\pi^{4}}\int Tr(F(\chi)T_{\chi}(\text{$\mathfrak{g}$}%
(x)^{-1})\chi\bar{\chi}d\chi
\]
and its inverse is
\[
F(\chi)=\int f(g)T_{\chi}(\mathfrak{g}(x))dx,
\]
where the $dx$ is the measure on $CS^{3}$. The measure $dx$ is equal to the
bi-invariant measure on $SL(2,C)$ under the isomorphism $\mathfrak{g}$.

The expansion of the delta function on $SL(2,C)$ from equation (\ref{invFour})
is%
\begin{equation}
\delta(g)=\frac{1}{8\pi^{4}}\int tr\left[  T_{\chi}(g)\right]  \chi\bar{\chi
}d\chi. \label{deltaExp}%
\end{equation}
Let me calculate the trace $tr\left[  T_{\chi}(g)\right]  $. If $\lambda
=e^{\rho+i\theta}$ and $\frac{1}{\lambda}$ are the eigen values of $g$ then%
\[
tr\left[  T_{\chi}(g)\right]  =\dfrac{\lambda^{\chi_{1}}\bar{\lambda}%
^{\chi_{2}}+\lambda^{-\chi_{1}}\bar{\lambda}^{-\chi_{2}}}{\left\vert
\lambda-\lambda^{-1}\right\vert ^{2}},
\]
which is to be understood in the sense of distributions \cite{IMG}. The trace
can be explicitly calculated as%
\begin{equation}
tr\left[  T_{\chi}(g)\right]  =\dfrac{\cos(\eta\rho+n\theta)}{2\left\vert
\sinh(\eta+i\theta)\right\vert ^{2}}. \label{eq.trsl(2,C)}%
\end{equation}
Therefore, the expression for the delta on $SL(2,C)$ explicitly is%
\begin{equation}
\delta(g)=\frac{1}{8\pi^{4}}\sum_{n}\int d\rho(n^{2}+\rho^{2})\dfrac{\cos
(\rho\eta+n\theta)}{\left\vert \sinh(\eta+i\theta)\right\vert ^{2}}.
\label{deltaExplicit}%
\end{equation}
Let us consider the integrand in equation (\ref{invFour}). Using equation
(\ref{Four}) in it we have
\begin{align}
Tr(F(\chi)T_{\chi}(g^{-1}))\chi\bar{\chi}  &  =\chi\bar{\chi}\int f(\acute
{g})Tr(T_{\chi}(\acute{g})T_{\chi}(g^{-1}))d\acute{g}\nonumber\\
&  =\chi\bar{\chi}\int f(\acute{g})Tr(T_{\chi}(\acute{g}g^{-1}))d\acute{g}.
\label{SignChi}%
\end{align}
But, since the trace is insensitive to an overall sign of $\chi$, so are the
terms of the Fourier expansion of the $L^{2}$ functions on $SL(2,C)\ $and
$CS^{3}$.

\section{Unitary Representations of $SO(4,C)$}

The group $SO(4,C)$ is related to its universal covering group $SL(2,C)\times
SL(2,C)$ by the relationship $SO(4,C)\approx\frac{SL(2,C)\times SL(2,C)}%
{Z^{2}}$. The map from $SO(4,C)$ to $SL(2,C)\times SL(2,C)$ is given by the
isomorphism between complex four vectors and $GL(2,C)$ matrices. If
$X=(a,b,c,d)$ then $G:C^{4}\longrightarrow GL(2,C)\ $can be defined by the
following:%
\[
G(X)=\left[
\begin{array}
[c]{cc}%
a+ib & c+id\\
-c+id & a-ib
\end{array}
\right]  .
\]
It can be easily inferred that $\det G(X)=a^{2}+$ $b^{2}+c^{2}+d^{2}$ is the
Euclidean norm of the vector $X$. Then, in general a $SO(4,C)$ rotation of a
vector $X$ to another vector $Y$ is given in terms of two arbitrary
$SL(2,C)\ $matrices $g_{L~B}^{~~A},~g_{R~B^{^{\prime}}}^{~~A^{^{\prime}}}\in
SL(2,C)$ by
\[
G(Y)^{AA^{^{\prime}}}=g_{L~B}^{~~A}g_{R~B^{^{\prime}}}^{~A^{^{\prime}}}%
G^{AB}(X),
\]
where $G^{AB}(X)$ is the matrix elements of $G(X)$. The above transformation
does not differentiate between $(L_{B}^{A},R_{B^{^{\prime}}}^{A^{^{\prime}}})$
and $(-L_{B}^{A},-R_{B^{^{\prime}}}^{A^{^{\prime}}})$ which is responsible for
the factor $Z_{2}$ in $SO(4,C)\approx\frac{SL(2,C)\times SL(2,C)}{Z^{2}}$.

The unitary representation theory of the group $SL(2,C)\times SL(2,C)$ is
easily obtained by taking the tensor products of two Gelfand-Naimarck
representations of $SL(2,C)$. The Fourier expansion for any function
$f(g_{L},g_{R})$ of the universal cover is given by%
\[
f(g_{L},g_{R})=\frac{1}{64\pi^{8}}\int\chi_{L}\bar{\chi}_{L}\chi_{R}\bar{\chi
}_{R}F(\chi_{L},\chi_{R})T_{\chi}(g_{L}^{-1})T_{\chi}(g_{R}^{-1})d\chi
_{L}d\chi_{R},
\]
where $\chi_{L}=\frac{n_{L}+i\rho_{L}}{2}$ and $\chi_{R}=\frac{n_{R}+i\rho
_{R}}{2}$. The Fourier expansion on $SO(4,C)$ is given by reducing the above
expansion such that $f(g_{L},g_{R})=f(-g_{L},-g_{R})$. From equation
(\ref{eq.trsl(2,C)}) I have%
\[
tr\left[  T_{\chi}(-g)\right]  =(-1)^{n}tr\left[  T_{\chi}(-g)\right]  ,
\]
where $\chi=\frac{n+i\rho}{2}$. Therefore%
\[
f(-g_{L},-g_{R})=\frac{1}{8\pi^{4}}\int\chi_{L}\bar{\chi}_{L}\chi_{R}\bar
{\chi}_{R}F(\chi_{L},\chi_{R})(-1)^{n_{L}+n_{R}}T_{\chi}(g_{L}^{-1})T_{\chi
}(g_{R}^{-1})d\chi_{L}d\chi_{R}.
\]
This implies that for $f(g_{L},g_{R})=f(-g_{L},-g_{R}),$ I must
have$~(-1)^{n_{L}+n_{R}}$ $=1$. From this, I can infer that the representation
theory of $SO(4,C)$ is deduced from the representation theory of
$SL(2,C)\times SL(2,C)$ by restricting $n_{L}+n_{R}$ to be even integers. This
means that $n_{L}$ and $n_{R}$ should be either both odd numbers or even
numbers. I would like to denote the pair $(\chi_{L},\chi_{R})$ ($n_{L}+n_{R}$
even) by $\omega$.

There are two Casimir operators available for $SO(4,C),$ namely $\varepsilon
_{IJKL}\hat{B}^{IJ}\hat{B}^{KL}$ and $\eta_{IK}\eta_{JL}\hat{B}^{IJ}\hat
{B}^{KL}$. The elements of the representation space $D_{\chi_{L}}\otimes$
$D_{\chi_{R}}$ are the eigen states of the Casimirs. On them, the operators
reduce to the following:
\begin{equation}
\varepsilon_{IJKL}\hat{B}^{IJ}\hat{B}^{KL}=\frac{\chi_{L}^{2}-\chi_{R}^{2}}%
{2}~~\text{and}%
\end{equation}%
\begin{equation}
\eta_{IK}\eta_{JL}\hat{B}^{IJ}\hat{B}^{KL}=\frac{\chi_{L}^{2}+\chi_{R}^{2}%
-2}{2}.
\end{equation}

\end{document}